%% file: main.tex
\documentclass[aps,prl,twocolumn,superscriptaddress,floatfix,nofootinbib]{revtex4-2}

\usepackage{amsmath,amssymb,bm,amsthm}
\usepackage{array}
\usepackage{graphicx}
\usepackage{xcolor}
\usepackage{multirow}
\usepackage{mathtools}
\usepackage{pifont}
\usepackage{longtable}
\usepackage[colorlinks=true,citecolor=blue,linkcolor=blue,urlcolor=blue,hypertexnames=false]{hyperref}

\newcommand{\bB}{\bm{B}}

\begin{document}

\input{main_body}

\clearpage
\setcounter{section}{0}
\setcounter{subsection}{0}
\setcounter{subsubsection}{0}
\setcounter{equation}{0}
\setcounter{figure}{0}
\setcounter{table}{0}
\renewcommand{\thesection}{\Roman{section}}
\renewcommand{\thesubsection}{\Alph{subsection}}
\renewcommand{\theequation}{\arabic{equation}}
\renewcommand{\thefigure}{S\arabic{figure}}
\renewcommand{\thetable}{S\arabic{table}}
\renewcommand{\theHsection}{supp.\arabic{section}}
\renewcommand{\theHsubsection}{supp.\arabic{section}.\arabic{subsection}}
\renewcommand{\theHequation}{supp.\arabic{equation}}
\renewcommand{\theHfigure}{supp.\arabic{figure}}
\renewcommand{\theHtable}{supp.\arabic{table}}

\begingroup
\centering
{\large\bfseries Supplemental Material: Magnetic-Field Selection of Magnetic Order in Altermagnets and Noncollinear Antiferromagnets\par}
\vspace{1.0em}
{Qiu-Shi Huang,$^{1}$ Chaoxi Cui,$^{1}$ Yilin Han,$^{1}$ Junxi Duan,$^{1}$ Zhi-Ming Yu,$^{1,*}$ and Yugui Yao$^{1}$\par}
\vspace{0.6em}
{\itshape $^{1}$Key Lab of advanced optoelectronic quantum architecture and measurement (MOE), Beijing Key Laboratory of Quantum Matter State Control and Ultra-Precision Measurement Technology, and School of Physics, Beijing Institute of Technology, Beijing 100081, China\par}
\vspace{0.6em}
{\small (Dated: \today)\par}
\vspace{0.4em}
{\small $^{*}$zhiming\_yu@bit.edu.cn\par}
\endgroup
\vspace{1.0em}

\input{supplement_body}

\end{document}

%% file: main_body.tex
\title{Magnetic-Field Selection of Magnetic Order in Altermagnets and Noncollinear Antiferromagnets}
\author{Qiu-Shi Huang}
\affiliation{Key Lab of advanced optoelectronic quantum architecture and measurement (MOE), Beijing Key Laboratory of Quantum Matter State Control and Ultra-Precision Measurement Technology, and School of Physics, Beijing Institute of Technology, Beijing 100081, China}

\author{Chaoxi Cui}
\affiliation{Key Lab of advanced optoelectronic quantum architecture and measurement (MOE), Beijing Key Laboratory of Quantum Matter State Control and Ultra-Precision Measurement Technology, and School of Physics, Beijing Institute of Technology, Beijing 100081, China}

\author{Yilin Han}
\affiliation{Key Lab of advanced optoelectronic quantum architecture and measurement (MOE), Beijing Key Laboratory of Quantum Matter State Control and Ultra-Precision Measurement Technology, and School of Physics, Beijing Institute of Technology, Beijing 100081, China}

\author{Junxi Duan}
\affiliation{Key Lab of advanced optoelectronic quantum architecture and measurement (MOE), Beijing Key Laboratory of Quantum Matter State Control and Ultra-Precision Measurement Technology, and School of Physics, Beijing Institute of Technology, Beijing 100081, China}

\author{Zhi-Ming Yu}
\email{zhiming\_yu@bit.edu.cn}
\affiliation{Key Lab of advanced optoelectronic quantum architecture and measurement (MOE), Beijing Key Laboratory of Quantum Matter State Control and Ultra-Precision Measurement Technology, and School of Physics, Beijing Institute of Technology, Beijing 100081, China}
\author{Yugui Yao}
\affiliation{Key Lab of advanced optoelectronic quantum architecture and measurement (MOE), Beijing Key Laboratory of Quantum Matter State Control and Ultra-Precision Measurement Technology, and School of Physics, Beijing Institute of Technology, Beijing 100081, China}

\date{\today}

\begin{abstract}
Conventional field selection of magnetic order relies on the Zeeman coupling, which however  vanishes in  the magnets without net magnetization---a rapidly growing class  including altermagnets (AMs), noncollinear antiferromagnets (nc-AFMs), and $\mathcal{PT}$-symmetric antiferromagnets ($\mathcal{PT}$-AFMs). 
Here we show that the quantity that fundamentally couples a magnet to  a uniform   magnetic field is not the magnetization, but the binary order parameter $\eta$ that labels the two time-reversal-related minima of the Landau free energy.
We develop a Landau theory of order selection based on $\eta$ under the constraints of magnetic point group (MPG) symmetry, in which $\eta$ couples to odd-degree  polynomials in the magnetic field.
Within this framework,  the linear term is the ferromagnetic Zeeman coupling, while higher-order couplings with leading degree $n=3,5,7,9$ naturally appear in AMs and nc-AFMs. In contrast, combined $\mathcal{PT}$ symmetry forbids any such coupling.
Consequently, it is the $(n-1)$-order magnetic susceptibility, rather than the net magnetization, that serves as the primary experimental observable to identify the magnetic order of AMs and nc-AFMs.
For all 122 MPGs, we classify the leading  coupling degree and the corresponding polynomial forms.
We demonstrate our framework in two representative materials: the AM MnF$_2$ and the nc-AFM MnTe$_2$.
We further construct a symmetry-allowed spin model  for an AM system to  reveal the microscopic origin of the higher-order coupling, and establish the coupling coefficient explicitly in terms of the spin-model parameters.
Our work  unifies the description of magnetic-order selection across magnets with and without net magnetization, offers a microscopic origin for this counterintuitive physics, and provides fingerprints for distinguishing intrinsic field selection from extrinsic switching.
\end{abstract}

\maketitle

\begin{figure}[t]
\centering
\includegraphics[width=\columnwidth]{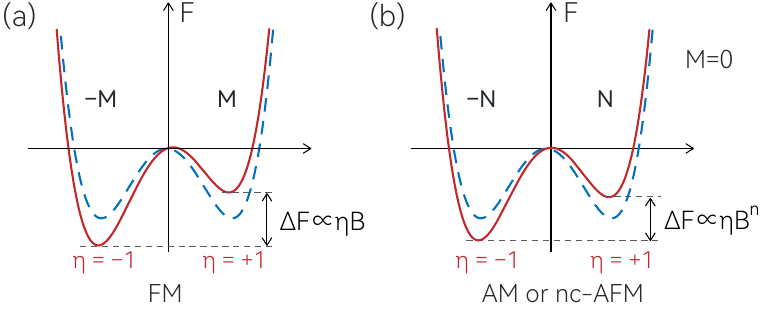}
\caption{\label{fig:double_well}
Illustration of $\bm B$ field induced  order selection for (a) FM and (b) AM and nc-AFM.
The binary variable $\eta=\pm 1$ defined by the two minima of Landau free energy is  a generic order label for order selection. 
$\eta$ couples linearly to $\bm B$ for a FM,  whereas  it couples to higher-order terms $B^n$ with odd $n\ge 3$ for an AM or nc-AFM. By contrast,   $\mathcal{PT}$ symmetry forbids any coupling between $\eta$ and $\bm B$.
The solid (dashed) curves denote the free energy in the presence (absence) of the $\bm B$ field.}
\end{figure}

{\emph{\textcolor{blue}{Introduction.--}}}
In the Landau theory of a magnetic transition, the free energy develops two degenerate, time-reversal ($\mathcal{T}$)-related minima below the critical temperature~\cite{LandauLifshitz1980}. These minima define a binary variable $\eta=\pm1$ that labels the two minima and thus the two opposite magnetic orders, as illustrated in Fig.~\ref{fig:double_well}. This binary degree of freedom and the ability to select one of them are the building blocks of magnetic memory and spintronics~\cite{LandauLifshitz1980,Chappert2007SpinElectronics,Marti2014AFMMemory,Kent2015MagneticMemories,Wadley2016CuMnAs,Jungwirth2016,Baltz2018,Zelezny2018SpinTorqueReview,Dieny2020Spintronics,Ma2021PiezomagneticAFM,Han2024NeelSwitching,He2024PerpendicularNeel,Zhou2025CrSbSymmetry,Zhou2026ChiralSwitching,Fu2026UnifiedSwitching}.

In ferromagnets (FMs), this selection is accomplished by the Zeeman coupling $-\bm M\cdot\bm B$, where the order parameter is the magnetization $\bm M$ itself. The Zeeman coupling tilts the double well and selects the minimum in which $\bm M$ is aligned with $\bm B$ [see Fig.~\ref{fig:double_well}(a)].
The paradigm, however, requires a net magnetization $\bm M$, and is therefore absent in the rapidly expanding class of fully compensated magnets---including $\mathcal{PT}$-symmetric antiferromagnets ($\mathcal{PT}$-AFMs)~\cite{Jungwirth2016,Baltz2018,Shao2021SpinNeutral,Shao2023NeelSpinCurrents,Shao2024AFMTJ,Tang2016PTAFM,Cao2023PTAHE,Liu2021CompensatedAHE,Feng2025DynamicSpin}, altermagnets (AMs)~\cite{Smejkal2022Altermagnetism,Smejkal2022Landscape,Bai2024Altermagnetism,PhysRevX.12.021016,McClarty2024Altermagnetism,Zhang2024GateAltermagnet,He2024Quasi1DTransport,Zhou2024CrystalThermal,Krempasky2024Kramers,Amin2024MnTeDomains,Duan2025AntiferroelectricAM,Zhu2026AltermagneticProximity,Liu2026AntiferroaxialAM,Huang2026LightOddParity,Zhu2026FloquetOddParity,Liu2026LightOddParityDimer,Li2026FloquetSpinSplitting,Zhang2026SwitchableOddParity,Li2026TunableFloquetBilayer,Zhou2025CrSbSymmetry}, and noncollinear antiferromagnets (nc-AFMs)~\cite{Feng2020NoncoplanarMO,Chen2014NoncollinearAHE,Kubler2014NoncollinearAHE,Nakatsuji2015Mn3Sn,Suzuki2017ClusterMultipole}.
For these systems, the Zeeman term vanishes, leading to the common belief that a uniform $\bm B$-field cannot intrinsically select between the two $\mathcal{T}$-related ordered states in fully compensated magnets~\cite{Jungwirth2016,Baltz2018}. Control strategies have consequently shifted to spin torques~\cite{Zelezny2014NeelTorque,Wadley2016CuMnAs,Grzybowski2017CuMnAs,Bodnar2018Mn2Au,Zelezny2018SpinTorqueReview,Baldrati2019NiO}, optical excitation~\cite{Nemec2018Opto,Higuchi2016MnF2,Kimel2009InertiaSwitching,Disa2020OpticalCrystalField,Afanasiev2021LightDrivenPhonons}, and magnetoelectric boundary effects~\cite{Fiebig2005Magnetoelectric,Belashchenko2010BoundaryME,Spaldin2008Toroidal,Fallarino2015ChromiaSwitching}.

However, several experiments have reported $\bm B$-field control of the magnetic-domain population in compensated magnets~\cite{Brown1998Cr2O3,Tardif2015Cd2Os2O7,Higuchi2016MnF2,Amin2024MnTeDomains,Hayashida2026DyFeO3}. The underlying mechanism remains poorly understood, and in some cases the switching has been attributed to extrinsic factors such as boundary effects, defects, or symmetry breaking~\cite{Belashchenko2010BoundaryME,Fallarino2015ChromiaSwitching}. 
Two fundamental questions therefore arise: can a uniform $\bm B$-field select the magnetic order of a magnet without net magnetization, and, if so, how can such intrinsic selection be distinguished from extrinsic switching?

In this work, we show that the answer to the first question is yes, and that the answer to the second follows from symmetry. The central observation is that the absence of $\bm M$ in compensated magnets does not eliminate the double well, suggesting that the binary variable $\eta$ is more fundamental than $\bm M$ for order selection.
Specifically, a uniform $\bm B$-field selects the magnetic order as long as  it can selectively tilt the double well.
We therefore develop a Landau theory of order selection based on $\eta$ under the constraints of magnetic point group (MPG) symmetry.
We find that $\eta$ can couple to odd-degree polynomials $\phi_i^n(\bm B)$ in $\bm B$, regardless of whether a net magnetization exists, and the form of $\phi_i^n(\bm B)$ is fully determined by the MPG.
Particularly, the symmetry-allowed form of $\phi_i^n(\bm B)$ provides fingerprints---the angular nodes of the field polynomials---for distinguishing intrinsic field selection from extrinsic switching.

By classifying the 122 MPGs according to the leading order of $\bm B$ in the coupling, we find three physical classes: FMs with a linear coupling, AMs and nc-AFMs with higher-order coupling, and $\mathcal{PT}$-AFMs that are $\bm B$-field silent without the  coupling.
This symmetry classification directly demonstrates that  a uniform $\bm B$ field can induce a tilt of the double well in the magnets both with and  without net magnetization.
We further demonstrate this framework in two representative materials: the AM  MnF$_2$ and the nc-AFM MnTe$_2$.
Moreover, we go beyond the symmetry classification, and construct a microscopic spin model that reveals the origin of the higher-order coupling.  

{\emph{\textcolor{blue}{Landau theory for magnetic-order selection.--}}}
Since the Landau double well exists for all the magnet, we develop a universal  framework of field-induced order selection based on the binary variable $\eta=\pm1$, which labels the two states of the magnet with opposite magnetic order.
In the presence of an external magnetic field $\bm B$, the Landau free energy of any magnet can be written in the general form
\begin{equation}
F = F_0 - \eta \sum_{i,n} \lambda_{i,n} \phi_i^{n}(\bm B)+\eta^2 \sum_{i,n} \lambda_{i,n}^{\prime} Q_i^{n}(\bm B),
\label{eq:free}
\end{equation}
where $F_0$ is the free energy of the system at zero field,  the last two terms respectively describe the coupling of the magnetic field to $\eta$ and $\eta^2$, $\lambda_{i,n}$ and $\lambda_{i,n}^{\prime}$ are material-dependent coupling coefficients, and $\phi_i^{n}$ and  $Q_i^{n}$ are the symmetry-allowed $n$th-order polynomials in  $\bm B$.

Clearly, only the second term in Eq. (\ref{eq:free}), which is linear in $\eta$, can tilt  the double well and then is relevant for order selection. We therefore only focus on this term.
Moreover, we expand the  Landau   free energy up to the leading order of $\bB$. 
Then, the relevant free energy for order selection  reduces to
\begin{equation}
{\cal F}= - \eta \sum_{i} \lambda_{i} \phi_i^{n}(\bm B),
\label{eq:free_energy_universal}
\end{equation}
with  $n$ here being a material-dependent constant.
The form of $\phi_i^{n}(\bm B)$ is fully determined by the symmetry of the magnets. 

The two ordered states of the magnets have same MPG, which is assumed as $\mathcal M$.  Since the two states are related by $\mathcal T$, $\mathcal M$ does not contain $\mathcal T$.  To investigate the field selection of magnetic order, we also should have a parent MPG $\mathcal G$, which describes the symmetry of the system before it enters into one of the two states by spontaneous symmetry breaking.  $\mathcal G$ is constructed as follows.  Define a group homomorphism $\pi: {\cal G} \to \mathbb{Z}_2$ by
\begin{equation}
{\cal M} \equiv \ker \pi = \{ g \in {\cal G} \mid \pi(g) = +1 \}.
\label{eq:domain_corep_kernel}
\end{equation}
Here, $\pi(g) = +1$ if $g$ preserves each magnetic order and $\pi(g) = -1$ if $g$ exchanges the two magnetic orders. 
Since the two states studied here are related by $\mathcal{T}$, the nontrivial coset of $\mathcal{M}$ in $\mathcal{G}$ is simply $\mathcal{M}\mathcal{T}$.  Consequently, the parent group $\mathcal G$ takes the form
\begin{equation}
\mathcal{G} = \mathcal{M} + \mathcal{M}\mathcal{T},
\end{equation}
which is a gray  MPG exhibiting $\mathcal T$ symmetry.

According to Landau theory~\cite{LandauLifshitz1980}, the free energy $\mathcal{F}$ in Eq. (\ref{eq:free_energy_universal}) must be invariant under every operator $g\in\mathcal G$. 
This symmetry constraint immediately leads to several remarkable physical consequences.

First, since $\mathcal{T}\eta = -\eta$ and $\mathcal{F}$ is $\mathcal{T}$-invariant, the invariance condition requires $\phi_i^{n}(-\bm B) = -\phi_i^{n}(\bm B)$. Hence $\phi_i^{n}(\bm B)$ must be an odd-degree polynomial in $\bm B$. That is, $n$ must be an odd number.

Second, if $\phi_i^{n}(\bm B)$ is linear in $\bm B$, the coupling term in Eq.~\eqref{eq:free_energy_universal} reduces to $-\bm M\cdot\bm B$ with $\bm M \equiv \eta \bm\lambda$, recovering the Zeeman coupling of a ferromagnet.
Alternatively, the linear coupling in $\mathcal{F}$ can be used to  define  ferromagnet, where  the magnetization is given as  $\bm M \equiv -\partial_{\bm B}\mathcal{F}$.
Consequently, magnets without net magnetization necessarily lack a linear coupling in $\mathcal{F}$.

Third, for $\mathcal{PT}$-AFMs, where $\mathcal P$ denotes spatial inversion,  their parent MPG $\mathcal G$ contains both $\mathcal P$ and $\mathcal T$. Since $\mathcal{PT}\eta = \eta$ and $\mathcal T\eta = -\eta$, it follows that $\mathcal P\eta = -\eta$. Because $\bm B$ is a pseudovector that is  invariant under $\mathcal P$, the $\mathcal{F}$ changes sign under $\mathcal P$ regardless of the degree of $\phi_i^n$. Consequently, $\mathcal{PT}$ symmetry enforces $\mathcal{F}(\eta,\bm B) = \mathcal{F}(\eta,-\bm B)$, forbidding any odd-degree field coupling and rendering field-induced magnetic-order selection impossible in $\mathcal{PT}$-AFMs. 
This might also partly explain the aforementioned common belief.

Fourth, AMs and nc-AFMs break $\mathcal{PT}$ symmetry, and may therefore exhibit  higher-order  coupling ($n \ge 3$) in $\mathcal{F}$, enabling magnetic-order selection by a uniform $\bB$ field. 
Moreover, the $\eta$-dependent higher-order  coupling  in free energy implies that the  higher-order magnetic susceptibility, defined as ${\chi}^{(n-1)} \equiv \partial^{n-1} {\bm M}/\partial {\bm B}^{n-1}= -\partial^{n} \mathcal{F}/\partial {\bm B}^{n}$, is also proportional to $\eta$, and thus takes opposite values for the two $\cal T$-related  ordered states. Therefore, in AMs and nc-AFMs, one can use the experimental observable ${\chi}^{(n-1)}$, instead of $\bm M$, to distinguish the two ordered states.

\begin{table*}[t]
\caption{\label{tab:classes}Classification of the 90 MPGs that accommodate magnetism. $n$ denotes the leading order of $\phi_i^n(\bB)$. The higher-order magnetic susceptibility ${\bm \chi}^{(n-1)}\propto \eta \lambda$ represents the primary experimental observable that can be  used  to distinguish the magnetic order of  AMs and  nc-AFMs.}
\footnotesize
\renewcommand{\arraystretch}{1.15}
\begin{ruledtabular}
\begin{tabular}{llll}
Magnetism & $n$ & MPGs & observable \\
\hline
FM & $1$ & $1$, $\bar{1}$, $2$, $2'$, $\mathrm m$, $\mathrm m'$, $2/\mathrm m$, $2'/\mathrm m'$, $2'2'2$, $\mathrm m'\mathrm m2'$, $\mathrm m'\mathrm m'2$, $\mathrm m'\mathrm m'\mathrm m$, $4$, $\bar{4}$, $4/\mathrm m$, $42'2'$, $4\mathrm m'\mathrm m'$ & ${\bm M}$ \\
& & $\bar{4}2'\mathrm m'$, $4/\mathrm m\mathrm m'\mathrm m'$, $3$, $\bar{3}$, $32'$, $3\mathrm m'$, $\bar{3}\mathrm m'$, $6$, $\bar{6}$, $6/\mathrm m$, $62'2'$, $6\mathrm m'\mathrm m'$, $\bar{6}\mathrm m'2'$, $6/\mathrm m\mathrm m'\mathrm m'$ \\
AM or & $3$ & $222$, $\mathrm{mm}2$, $\mathrm{mmm}$, $4'$, $\bar{4}'$, $4'/\mathrm m$, $4'22'$, $4'\mathrm m'\mathrm m$, $\bar{4}'2'\mathrm m$, $\bar{4}'2\mathrm m'$, $4'/\mathrm m\mathrm m'\mathrm m$, $32$, $3\mathrm m$, $\bar{3}\mathrm m$, $6'$ & ${\bm \chi}^{(2)}$ \\
nc-AFM & & $\bar{6}'$, $6'/\mathrm m'$, $6'22'$, $6'\mathrm m\mathrm m'$, $\bar{6}'\mathrm m'2$, $\bar{6}'\mathrm m2'$, $6'/\mathrm m'\mathrm{mm}'$, $23$, $\mathrm m\bar{3}$, $4'32'$, $\bar{4}'3\mathrm m'$, $\mathrm m\bar{3}\mathrm m'$ \\
& $5$ & $422$, $4\mathrm{mm}$, $\bar{4}2\mathrm m$, $4/\mathrm{mmm}$ & ${\bm \chi}^{(4)}$ \\
& $7$ & $622$, $6\mathrm{mm}$, $\bar{6}\mathrm m2$, $6/\mathrm{mmm}$ & ${\bm \chi}^{(6)}$ \\
& $9$ & $432$, $\bar{4}3\mathrm m$, $\mathrm m\bar{3}\mathrm m$ & ${\bm \chi}^{(8)}$ \\
$\mathcal{PT}$-AFM & \ding{55} & $\bar{1}'$, $2'/\mathrm m$, $2/\mathrm m'$, $\mathrm m'\mathrm{mm}$, $\mathrm m'\mathrm m'\mathrm m'$, $4/\mathrm m'$, $4'/\mathrm m'$, $4/\mathrm m'\mathrm{mm}$, $4'/\mathrm m'\mathrm m'\mathrm m$, $4/\mathrm m'\mathrm m'\mathrm m'$, $\bar{3}'$, $\bar{3}'\mathrm m$ & \ding{55} \\
& & $\bar{3}'\mathrm m'$, $6'/\mathrm m$, $6/\mathrm m'$, $6/\mathrm m'\mathrm{mm}$, $6'/\mathrm{mmm}'$, $6/\mathrm m'\mathrm m'\mathrm m'$, $\mathrm m'\bar{3}'$, $\mathrm m'\bar{3}'\mathrm m$, $\mathrm m'\bar{3}'\mathrm m'$ \\
\end{tabular}
\end{ruledtabular}
\end{table*}

We now investigate the explicit expression of $\phi_i^{n}(\bm B)$ through the corepresentations (coreps) of $\cal G$~\cite{BradleyCracknell1972}.
$\phi_i^n(\bm B)$  is constructed by the  monomials in the field components. Throughout, $B_x$, $B_y$, and $B_z$ denote components in an orthonormal Cartesian frame; for trigonal and hexagonal settings, $x$ is chosen along the first conventional basal axis and $z$ along the principal axis.
The set of all degree-$n$ monomials in  ${\bm B}$, 
\begin{equation}
\mathbf{V}^{(n)} = \{B_x^p B_y^q B_z^r \mid p+q+r=n\},
\end{equation}
forms a basis with dimension $d_n = (n+1)(n+2)/2$, which  generally corresponds to  a reducible $d_n$ dimensional correp $\Gamma_{n}^B$ of ${\cal G}$. 
Besides, we assume the 1D real  corep of  $\eta$ is $\Gamma_{\eta}$, which is 1 for the  operator  $g\in\mathcal M$ and is $-1$ for  $g\in\mathcal {MT}$.
A  coupling between $\eta$ and a $n$th-degree field polynomial is enabled, if and only if the direct product $\Gamma_{\eta}\otimes\Gamma_{n}^B$ contains the 1D trivial corep $\Gamma_1$ of ${\cal G}$~\cite{LandauLifshitz1980,McClarty2024Altermagnetism}:
\begin{equation}\label{eq:landau_irrep_product_0}
\Gamma_1\subset \Gamma_{\eta}\otimes\Gamma_{n}^B.
\end{equation}
Since $\Gamma_{\eta}$ is a real 1D corep, this condition is equivalent to $\Gamma_{\eta}\subset\Gamma_{n}^B$, i.e., the  correp $\Gamma_{n}^B$ must contain $\Gamma_{\eta}$ in its decomposition.

When Eq.~\eqref{eq:landau_irrep_product_0} is satisfied, the symmetry-allowed polynomials $\phi_i^n(\bB)$ can be extracted from the  ${\mathbf{V}}^{(n)}$ space via the projection operator onto the $\Gamma_{\eta}$ corep of ${\cal G}$:
\begin{equation}
\hat{P}^{\Gamma_\eta} = \frac{1}{|{\cal G}|} \sum_{g \in {\cal M}}  \hat{R}(g)-\frac{1}{|{\cal G}|} \sum_{g \in {\cal MT}}  \hat{R}(g),
\end{equation}
where  $|{\cal G}|$ is the order of the group ${\cal G}$,  and $\hat{R}(g)$ is the operator representing $g$ on the  monomial in ${\mathbf{V}}^{(n)}$.
Applying $\hat{P}^{\Gamma_\eta}$ to each  monomial in $\mathbf{V}^{(n)}$ and collecting the non-vanishing results yields the set  of symmetry-allowed $\phi_i^n(\bB)$. One can check that the resulting $\phi_i^n(\bB)$  transform as $\Gamma_{\eta}$.

Based on the above process, we investigate the explicit expression of $\phi_i^{n}(\bm B)$ for all 122 MPGs, and find that 31 MPGs allow a linear coupling, 38 allow a higher-order coupling, and 21 have ${\cal PT}$ symmetry, as listed in Table~\ref{tab:classes}. The complete leading-order basis is given in the Supplemental Material~\cite{SupplementalMaterial}\nocite{BilbaoCrystallographicServer,Aroyo2006BCS,Aroyo2006BCSII,BorovikRomanov2013,Elcoro2021MagTQC,Xu2020MagTopological,Dzyaloshinsky1958,Moriya1960,AlteDaVeiga1982MnF2,MommaIzumi2011VESTA}. Representative materials and their $\phi_i^{n}(\bm B)$ are summarized in Table~\ref{tab:materials_selectors}.

Next, we illustrate this framework with two representative examples.

\begin{figure}[b]
\centering
\includegraphics[width=\columnwidth]{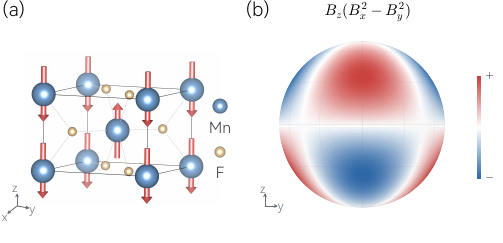}
\caption{\label{fig:mnf2_selector}
(a) Crystalline structure and magnetic configuration  of  MnF$_2$. 
(b) Free energy splitting $\Delta F$ between the ordered states with $\eta=\pm1$  as a function of $\bm B/B$, calculated from Eq. (\ref{eq:mnf2_free_energy}). 
Here, we set $\lambda_{\rm MnF_2}=3\sqrt{3}/4~\mathrm{meV\,T^{-3}}$ and $B=1~\mathrm{T}$.}
\end{figure}

{\emph{\textcolor{blue}{AM MnF$_2$.--}}}
Bulk rutile MnF$_2$ is a collinear AM belonging to MPG $\mathcal M = 4'/mm'm$~\cite{Higuchi2016MnF2,Bhowal2024Octupoles,Morano2025MnF2,Lovesey2026MnF2}, and then the corresponding parent MPG is  \(\mathcal G=\mathcal M+\mathcal M\mathcal T=4/mmm1'\). MnF$_2$ crystallizes in a tetragonal structure with lattice parameters $a=4.8734$~\AA{} and $c=3.3099$~\AA{} below N\'eel temperature $T_N=67.7$ K~\cite{VanHaren2023MnF2,Sun2025MnF2Exchange,Morano2025MnF2}. The local moments reside primarily on the Mn sites, with an easy axis along the $c$ direction, as illustrated in Fig.~\ref{fig:mnf2_selector}(a).

The generators of $\mathcal M$ can be chosen as \(C_{4z}\mathcal T\), \(C_{2,[110]}\), and \(\mathcal P\). 
Since $\eta$ is invariant under the generators of $\mathcal M$ but changes sign under $\mathcal T$, it belongs to the 1D correp $mB_{2g}$ of  $\mathcal G$ (see Table~\ref{tab:mnf2_selected_coreps} in the End Matter). 
According to Table~\ref{tab:mnf2_selected_coreps}, the linear basis $\mathbf{V}^{(1)} = \{B_x, B_y, B_z\}$ forms a reducible 3D corep that decomposes as $mA_{2g}\oplus mE_g$.  Thus, a linear coupling between $\eta$ and $\boldsymbol{B}$ is symmetry-forbidden in MnF$_2$, consistent with its altermagnetic character.

We then apply the projection operator $\hat{P}^{\Gamma_\eta}$ to the basis vectors in $\mathbf{V}^{(3)}$ to obtain a  basis that transforms  as  $\Gamma_\eta=mB_{2g}$, and find that  only polynomial $B_z(B_x^2 - B_y^2)$ persists (see Table~\ref{tab:mnf2_selected_coreps}). The free energy consequently takes the form
\begin{equation}
{\cal F}_{\rm MnF_2} = -\eta\lambda_{\rm MnF_2} B_z(B_x^2 - B_y^2),
\label{eq:mnf2_free_energy}
\end{equation}
indicating that the energy splitting between the two ordered states ($\eta = \pm 1$) is very sensitive to the orientation of $\boldsymbol{B}$. The splitting is maximal when $\boldsymbol{B}$ lies along the $(011)$ or $(101)$ direction, and vanishes when $B_z = 0$ or $|B_x| = |B_y|$, as shown in Fig.~\ref{fig:mnf2_selector}(b).

This cubic coupling further implies several nonvanishing $\eta$-dependent second-order susceptibilities $\chi_{ijk}^{(2)} \equiv \partial^2 M_i / (\partial B_j \partial B_k)$, such as  $\chi_{zxx}^{(2)} = -\chi_{zyy}^{(2)} = 2\eta\lambda_{\rm MnF_2}$. The coefficient $\lambda_{\rm MnF_2}$ can therefore be extracted experimentally, and the susceptibilities $\chi_{ijk}^{(2)}$ themselves provide a practical mean to distinguish the two $\mathcal T$-related magnetic orders of MnF$_2$.

\begin{figure}[t]
\centering
\includegraphics[width=\columnwidth]{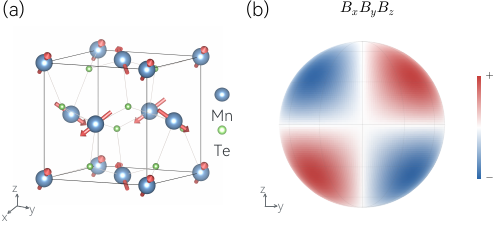}
\caption{\label{fig:mnte2_selector}
(a) Crystalline structure and magnetic configuration  of  MnTe$_2$. 
(b) Free energy splitting $\Delta F$ between the ordered states with $\eta=\pm1$  as a function of $\bm B/B$, calculated from Eq. (\ref{FreeE_Mnte}). 
Here, we set $\lambda_{\rm MnTe_2}=3\sqrt{3}/2~\mathrm{meV\,T^{-3}}$ and $B=1~\mathrm{T}$.}
\end{figure}

{\emph{\textcolor{blue}{nc-AFM MnTe$_2$.--}}}
Bulk MnTe$_2$ has a noncollinear all-in-all-out magnetic configuration, belonging to MPG \(\mathcal M=m\bar{3}\), as shown in Fig.~\ref{fig:mnte2_selector}(a)~\cite{Burlet1997MnTe2,Fahmy2025MnTe2,Palasyuk2026MnTe2,Feng2026MnTe2}.
Neutron diffraction gives a cubic lattice parameter $a\simeq6.9$~\AA{} and a magnetic transition temperature $T_N=86.5$~K~\cite{Burlet1997MnTe2}.
The corresponding parent group for MnTe$_2$ is \(\mathcal G=m\bar{3}1'\).

From Table~\ref{tab:mnte2_selected_coreps}, the variable \(\eta\) belongs to the 1D correp $mA_g$, while the linear basis $\mathbf{V}^{(1)} = \{B_x, B_y, B_z\}$ belongs to 3D irreducible correp $mT_g$.  Their distinct correps forbid a linear coupling between \(\eta\) and $\bB$ in MnTe$_2$.
Similarly, via projection operator, we find that only the monomial $B_xB_yB_z$ is symmetry allowed, and the relevant free energy is 
\begin{equation}
{\cal F}_{\rm MnTe_2}=-\eta\lambda_{\rm MnTe_2} B_xB_yB_z.
\label{FreeE_Mnte}
\end{equation}
Here, the splitting is maximal when $\boldsymbol{B}$ lies along the $(111)$  direction, and vanishes when $B_x = 0$, $B_y = 0$ or $B_z=0$, as shown in Fig.~\ref{fig:mnte2_selector}(b).
All the six symmetry-allowed second-order susceptibilities of MnTe$_2$ are identical \(\chi_{ijk}^{(2)}=\eta\lambda_{\rm MnTe_2}\) with $\{i,j,k\}\in\{x,y,z\}$ and $i\neq j \neq k$.

\begin{figure}[h!]
\centering
\includegraphics[width=\columnwidth]{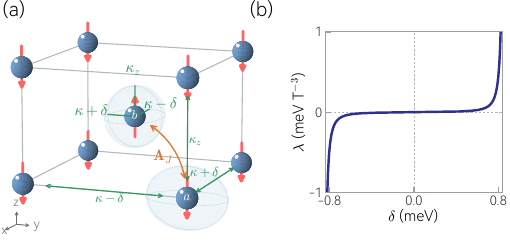}
\caption{\label{fig:lambda_delta}
(a) Illustration of the spin lattice model (\ref{eq:mnf2_uniform_spin_hamiltonian}), where only the magnetic atoms are plotted. The ellipsoids show that the surroundings  for the  two  magnetic sublattices  are different.
(b) The coupling coefficient $\lambda$ vs the $\delta$.
Here, we set $K=1~\mathrm{meV}$, $J_\perp=1~\mathrm{meV}$, and $h=0.3~\mathrm{meV\,T^{-1}}$.}
\end{figure}

{\emph{\textcolor{blue}{Microscopic origin.--}}}
To uncover the microscopic origin of the nonlinear coupling, we construct a minimal two-sublattice AM spin model belonging to MPG $4'/mm'm$.
Because MPG $4'/mm'm$ lacks $\mathcal{PT}$ symmetry, the atomic environments of the two magnetic  sublattices are different, as indicated by the ellipsoids in Fig.~\ref{fig:lambda_delta}(a).
Let $\bm S_a$ and $\bm S_b$ denote the normalized magnetic  moment  of the two magnetic sublattices, with $\bm S_a=-\bm S_b=\eta\hat{\bm z}$ at zero field [see Fig.~\ref{fig:lambda_delta}(a)].
The symmetry-allowed spin Hamiltonian of the system is~\cite{SupplementalMaterial}
\begin{equation}
H_0=\sum_i \left({\bm S}_{i,a}^{\mathsf T}\boldsymbol{A}_{\kappa}^a\bm S_{i,a}+\bm S_{i,b}^{\mathsf T}\boldsymbol{A}_{\kappa}^b\bm S_{i,b}\right)
+\sum_{\langle i,j\rangle}\bm S_{i,a}^{\mathsf T}\boldsymbol{A}_{J}\bm S_{j,b},
\label{eq:mnf2_uniform_spin_hamiltonian}
\end{equation}
where the first sum describes the anisotropy of the two  magnetic sublattices, and $\langle i,j\rangle$ in the second term denotes nearest-neighbor coupling.
The non-vanishing components of the tensors $\boldsymbol{A}_{\kappa}^{a(b)}$ and $\boldsymbol{A}_{J}$ are
$A_{\kappa,xx}^a = A_{\kappa,yy}^b = \kappa_x$,
$A_{\kappa,yy}^a = A_{\kappa,xx}^b = \kappa_y$,
$A_{\kappa,zz}^{a(b)} = \kappa_z$,
$A_{J,xx} = A_{J,yy} = J_{\perp}$,
and $A_{J,zz} = J_z$.
If the  $\mathcal{PT}$ symmetry is  imposed, the environments of the two magnetic sublattices would become identical, enforcing $\kappa_x = \kappa_y$.
It is therefore natural to introduce
$\delta \equiv (\kappa_x - \kappa_y)/2$
as the parameter that measures $\mathcal{PT}$-symmetry breaking.

In the presence of a magnetic field, the free energy per magnetic unit cell reads
\begin{equation}
F=U_0-h\bm B\cdot(\bm S_a+\bm S_b),
\label{eq:mnf2_field_free_energy}
\end{equation}
where $h$ denotes the magnitude of the magnetic moment, and
\begin{eqnarray}\label{eq:mnf2_full_spin_hamiltonian}
U_0&=& \kappa \left(\rho_{a,+}+\rho_{b,+}\right)+\delta\left(\rho_{a,-}-\rho_{b,-}\right) \nonumber \\ 
&&+\kappa_z\left(S_{a,z}^2+S_{b,z}^2\right)+J_\perp\left(S_{a,x}S_{b,x}+S_{a,y}S_{b,y}\right)\nonumber\\
&&+J_zS_{a,z}S_{b,z},
\end{eqnarray}
with $\kappa = (\kappa_x + \kappa_y)/2$ and $\rho_{a(b),\pm} = S_{a(b),x}^2 \pm S_{a(b),y}^2$.

Since the moment magnitude   $h$ generally is  unaffected by a moderate field, only the spin orientations $\bm S_{a}$ and $\bm S_{b}$ vary.
The variation of $\bm S_{a(b)}$ is determined  by the stationary condition of the free energy [Eq.~\eqref{eq:mnf2_field_free_energy}] with respect to the spin orientations.
A straightforward stationary-point calculation~\cite{SupplementalMaterial} gives
\begin{equation}
\begin{aligned}
\bm S_a={}&\left(\alpha B_x,\,\beta B_y,\,
\eta\sqrt{1-\alpha^2B_x^2-\beta^2B_y^2}\right),\\
\bm S_b={}&\left(\beta B_x,\,\alpha B_y,\,
-\eta\sqrt{1-\beta^2B_x^2-\alpha^2B_y^2}\right),
\end{aligned}
\label{eq:mnf2_canting_main}
\end{equation}
where $\alpha=h(2K-2\delta-J_\perp)/(4K^2-4\delta^2-J_\perp^2)$ and $\beta=h(2K+2\delta-J_\perp)/(4K^2-4\delta^2-J_\perp^2)$, with $K=\kappa-\kappa_z+J_z/2$.
The in-plane canting is linear in $\bm B$, whereas the corrections to the $z$ components are of second order.
According to the $2n+1$ theorem~\cite{GonzeVigneron1989,Solovyev2014SCLR}, the first-order response [Eq.~\eqref{eq:mnf2_canting_main}] suffices to determine the free energy correctly up to third order in $\bm B$.

Substituting Eq.~\eqref{eq:mnf2_canting_main} into Eq.~\eqref{eq:mnf2_field_free_energy}, one finds that, among all generated terms, only $-hB_z(S_{a,z}+S_{b,z})$ is linear in $\eta$ and odd in $\bm B$, which is therefore solely responsible for the field-induced tilt of the double well.
Notice that for $\delta=0$, corresponding to the system with ${\cal PT}$ symmetry,  one has $\alpha=\beta$, i.e. the two sublattices cant identically.
This makes $S_{a,z}+S_{b,z}=0$ and then forbids  the higher-order coupling.
In contrast,  $\delta\neq0$  renders the field responses of the two sublattices unequal, leading to  a third-order coupling.

Explicitly, the free energy $F$~\eqref{eq:mnf2_field_free_energy} up to third order in $\bm B$ is
\begin{equation}
\begin{aligned}
F(\bm B)={}&F_0 + \lambda^{\prime}(B_x^2+B_y^2)-\eta\lambda B_z(B_x^2-B_y^2),
\end{aligned}
\label{eq:mnf2_microscopic_free_energy_main}
\end{equation}
where $F_0$ is the free energy at zero field,
$\lambda^{\prime}=-{h^2(2K-J_\perp)}/{(4K^2-4\delta^2-J_\perp^2)}$.
The resulting nonlinear coupling coefficient is
\begin{equation}
\lambda=\frac{h}{2}(\beta^2-\alpha^2)
=\frac{4h^3(2K-J_\perp)\,\delta}
{(4K^2-4\delta^2-J_\perp^2)^2}.
\label{eq:mnf2_lambda}
\end{equation}
The third-order term in Eq.~\eqref{eq:mnf2_microscopic_free_energy_main} exactly matches the symmetry-determined form of Eq.~\eqref{eq:mnf2_free_energy}.
Importantly, while the coupling is induced by $\mathcal{PT}$ breaking---$\lambda$ is odd in $\delta$ and vanishes for   $\delta=0$---it is not simply linear in $\delta$.
Particularly, $\lambda$ may be significantly enhanced when $\delta$ approaches  the  critical value $\delta_c=\sqrt{4K^2-J_\perp^2}/2$, as illustrated in Fig.~\ref{fig:lambda_delta}(b).

{\emph{\textcolor{blue}{Conclusions.--}}}
We have established a general symmetry framework for magnetic-order selection by a uniform magnetic field, based on  $\eta$ that labels the two ${\cal T}$-related minima of the Landau free energy.
We find that  a uniform field can intrinsically select between ${\cal T}$-related magnetic orders  even in fully compensated magnets, and the $\eta$-odd higher-order susceptibility $\bm\chi^{(n-1)}$ can be used  in distinguishing the two  magnetic orders.

Moreover, the field-induced magnetization also exhibit many interesting  features. 
For example, in MnF$_2$,  a field applied along $x$ can generate a $\eta$-dependent  $z$-direction magnetization, as  $M_z=-\partial_{B_z}F=\eta\lambda_{\rm MnF_2} B_x^2$, and in MnTe$_2$ a field along  the $(110)$ direction  induces $M_z=\eta\lambda_{\rm MnTe_2} B_xB_y$.
This  means that  the field-induced magnetization  for AM and nc-AFM, which  scale as $B^2$ and is $\eta$-odd,  can also detect the magnetic order of system.

{\emph{Acknowledgments.--}} This work is supported by the National Key R\&D Program of China (Grant No.~2025YFA1411200) and the National Natural Science Foundation of China (Grant Nos.~12234003, 12474040, 12474039, and 125B2096).

\input{main.bbl}
\onecolumngrid 
\appendix     

\section*{End Matter} 

{\emph{\textcolor{blue}{Corep of  MPGs $4/mmm1'$ and  $m\bar{3}1'$.--}}}
We present part of the  single-valued corep of  MPG $4/mmm1'$  in Table \ref{tab:mnf2_selected_coreps}, and that of  MPG  $m\bar{3}1'$ in Table \ref{tab:mnte2_selected_coreps}.
The generators of $m\bar{3}$ can be chosen as $C_{2z}$,  $C_{3,[111]}$, and  $\mathcal P$.
Besides,  MPG $m\bar{3}1'=m\bar{3}\times\{E,{\cal T}\}$.

\vspace{0.7em}
\noindent
\begin{minipage}[t]{0.49\textwidth}
\vspace{0pt}
\refstepcounter{table}\label{tab:mnf2_selected_coreps}
\parbox[t][3\baselineskip][t]{\linewidth}{TABLE~\Roman{table}. Part of the single-valued corep of MPG $4/mmm1'$, including the generators of both $4'/mm'm$ and $4/mmm1'$~\cite{Hlinka2025MagneticAizu}.}
\par\vspace{0.35em}
\footnotesize
\renewcommand{\arraystretch}{1.15}
\begin{ruledtabular}
\begin{tabular}{lcccc l}
Corep & $C_{4z}\mathcal T$ & $C_{2,[110]}$ & $\mathcal P$ & $\mathcal T$ & Basis \\
\hline
$A_{1g}$ ($\Gamma_1$)  & 1  & 1  & 1 & 1  & $1$, $B_x^2+B_y^2$, $B_z^2$ \\
$mA_{2g}$ & -1 & -1 & 1 & -1 & $B_z$ \\
$mB_{2g}$ & 1  & 1  & 1 & -1 & $\eta$, $B_z(B_x^2-B_y^2)$ \\
$mE_g$    & 0  & 0  & 2 & -2 & $(B_x,B_y)$
\end{tabular}
\end{ruledtabular}
\end{minipage}
\hfill
\begin{minipage}[t]{0.49\textwidth}
\vspace{0pt}
\refstepcounter{table}\label{tab:mnte2_selected_coreps}
\parbox[t][3\baselineskip][t]{\linewidth}{TABLE~\Roman{table}. Part of the single-valued corep of MPG $m\bar{3}1'$, including the generators of both $m\bar{3}$ and $m\bar{3}1'$~\cite{Hlinka2025MagneticAizu}.}
\par\vspace{0.35em}
\footnotesize
\renewcommand{\arraystretch}{1.3}
\begin{ruledtabular}
\begin{tabular}{l c c c c l}
Corep & $C_{2z}$ & $C_{3,[111]}$ & $\mathcal P$ & $\mathcal T$ & Basis \\
\hline
$A_g$ ($\Gamma_1$)  & 1  & 1 & 1 & 1  & $1$, $B_x^2+B_y^2+B_z^2$ \\
$mA_g$  & 1  & 1 & 1 & -1 & $\eta$, $B_xB_yB_z$ \\
$mT_g$ & -1 & 0 & 3 & -3 & $(B_x,B_y,B_z)$
\end{tabular}
\end{ruledtabular}
\end{minipage}

\begin{table}[h!]
\caption{\label{tab:materials_selectors}Representative experimentally synthesized AM and nc-AFM materials.}
\footnotesize
\renewcommand{\arraystretch}{1.15}
\begin{ruledtabular}
\begin{tabular}{lclcl}
Magnetism & $n$ & Materials & MPG & $\phi_i^n(\bB)$ \\
\hline
AM & $3$ & MnF$_2$~\cite{VanHaren2023MnF2} & $4'/\mathrm m\mathrm m'\mathrm m$ & $B_z(B_x^2-B_y^2)$ \\
& & FeS~\cite{Takagi2025FeS,Wang2026TypeIIAFE} & $\bar{6}'\mathrm m'2$ & $B_y(3B_x^2-B_y^2)$ \\
& & MnTe~\cite{Amin2024MnTeDomains,Krempasky2024Kramers,Kriegner2016MnTeMemory,Lee2024MnTeKramers,Hariki2024MnTeXMCD,Liu2024MnTeChiralMagnon,Takegami2025MnTeDomains,Chen2025MnTeElectricalSwitching,Dong2026MnTePhotocurrent,Liu2026MnTeChiralMagnons,Liu2026MnTeOrderSwitching,Yang2026MnTeMagnetoOptics} & $\mathrm{mmm}$ & $B_xB_yB_z$ \\
& & DyFeO$_3$ ($\Gamma_1$ phase)~\cite{Hayashida2026DyFeO3} & $\mathrm{mmm}$ & $B_xB_yB_z$ \\
& & CrSb~\cite{Yuan2020CrSb,Reimers2024CrSb,Zhou2025CrSbSymmetry} & $6'/\mathrm m'\mathrm{mm}'$ & $B_x(B_x^2-3B_y^2)$ \\
& $5$ & Ba$_2$MnSi$_2$O$_7$~\cite{Sale2019Ba2MnSi2O7} & $\bar{4}2\mathrm m$ & $B_xB_yB_z(B_x^2-B_y^2)$ \\
& & KMnF$_3$ (tetragonal AFM phase)~\cite{Knight2020KMnF3} & $4/\mathrm{mmm}$ & $B_xB_yB_z(B_x^2-B_y^2)$ \\
nc-AFM & $3$ & MnTe$_2$~\cite{Burlet1997MnTe2,Feng2026MnTe2} & $\mathrm m\bar{3}$ & $B_xB_yB_z$ \\
& & Cd$_2$Os$_2$O$_7$~\cite{Tardif2015Cd2Os2O7} & $\mathrm m\bar{3}\mathrm m'$ & $B_xB_yB_z$ \\
& $5$ & Ho$_2$Ge$_2$O$_7$~\cite{Morosan2008Ho2Ge2O7} & $422$ & $B_xB_yB_z(B_x^2-B_y^2)$ \\
& $7$ & Ba$_3$CoSb$_2$O$_9$~\cite{Ma2016Ba3CoSb2O9} & $\bar{6}\mathrm m2$ & $B_xB_yB_z(B_x^2-3B_y^2)(3B_x^2-B_y^2)$ \\
& $9$ & SrCuTe$_2$O$_6$~\cite{Chillal2020SrCuTe2O6} & $432$ & $B_xB_yB_z(B_x^2-B_y^2)(B_x^2-B_z^2)(B_y^2-B_z^2)$ \\
\end{tabular}
\end{ruledtabular}
\end{table}

%% file: supplement_body.tex
\section{Magnetic-point-group scan}

The 122 crystallographic magnetic point groups (MPGs) comprise 32 gray groups that contain pure time reversal and 90 non-gray groups~\cite{SM:BilbaoCrystallographicServer,SM:Aroyo2006BCS,SM:BorovikRomanov2013}.  A single member of a time-reversed magnetic-domain pair cannot contain pure \(\mathcal T\); its ordered-state MPG \(\mathcal M\) therefore belongs to one of the 90 non-gray classes.  The corresponding symmetry before spontaneous domain selection is the gray parent
\begin{equation}
\mathcal G=\mathcal M+\mathcal M\mathcal T.
\label{eq:gray_parent_supp}
\end{equation}
Equivalently, the two cosets define a real one-dimensional domain character
\begin{equation}
\pi(g)=
\begin{cases}
+1,&g\in\mathcal M,\\
-1,&g\in\mathcal M\mathcal T,
\end{cases}
\qquad
\mathcal M=\ker\pi,
\qquad
\pi(\mathcal T)=-1.
\label{eq:domain_character_supp}
\end{equation}
Thus the 32 gray classes serve as parent groups, whereas the physical classification in the main text is indexed by the 90 possible ordered-state subgroups \(\mathcal M\).

Throughout, \(B_x,B_y,B_z\) are components in an orthonormal Cartesian frame. For trigonal and hexagonal settings, \(x\) is chosen along the first conventional basal axis and \(z\) along the principal axis, and \(\boldsymbol R_g\) below is the corresponding orthogonal Cartesian matrix.

For a magnetic operation \(g=(\boldsymbol R_g,\epsilon_g)\), where \(\epsilon_g=0\) for a unitary operation and \(1\) for an antiunitary operation, a uniform magnetic field transforms as the axial, time-reversal-odd vector
\begin{equation}
\bB\longmapsto \boldsymbol D_B(g)\bB,
\qquad
\boldsymbol D_B(g)=(-1)^{\epsilon_g}\det(\boldsymbol R_g)\boldsymbol R_g.
\label{eq:field_action_supp}
\end{equation}
A degree-\(n\) polynomial \(\phi_i^n(\bB)\) can couple linearly to the domain variable only when it transforms with the same character,
\begin{equation}
\phi_i^n\!\left[\boldsymbol D_B(g)\bB\right]=\pi(g)\phi_i^n(\bB),
\qquad g\in\mathcal G.
\label{eq:selector_covariance_supp}
\end{equation}
To solve Eq.~\eqref{eq:selector_covariance_supp}, we represent the monomial basis
\begin{equation}
\mathbf V^{(n)}=\{B_x^pB_y^qB_z^r\mid p+q+r=n\}
\end{equation}
by matrices \(\boldsymbol T_n(g)\).  If \(\bm c\) is the coefficient vector of a polynomial in this basis, every allowed selector lies in the common null space
\begin{equation}
\left[\boldsymbol T_n(g)-\pi(g)\boldsymbol I\right]\bm c=0,
\qquad g\in\mathcal G.
\label{eq:selector_nullspace_supp}
\end{equation}
Because \(\pi(\mathcal T)=-1\), only odd \(n\) need be considered.

Applying this construction to the 90 non-gray MPGs gives 31 groups with a linear selector and 38 compensated \(\mathcal{PT}\)-breaking groups whose leading selector occurs at \(n=3,5,7\), or 9.  For the remaining 21 groups, \(\bar{1}'=\mathcal{PT}\) belongs to \(\mathcal M\).  Since an axial field is invariant under inversion while the domain variable is odd under inversion, Eq.~\eqref{eq:selector_covariance_supp} has no nonzero odd pure-field solution at any order.  These groups are therefore exactly pure-field silent, rather than silent only up to a finite search rank.  A standard MPG symbol alone does not retain how the subgroup is embedded in its gray parent.  Table~\ref{tab:mpg-character-scan} therefore gives the complete channel-resolved result for all 105 time-reversal-odd one-dimensional corepresentation channels of the 32 gray parents, listing the ordered-state subgroup \(\mathcal M\), its parent \(\mathcal G\), the lowest allowed odd rank, the physical class, and explicit bases of \(\phi_i^n(\bB)\), cross-checked against the BCS COREPRESENTATIONS PG data~\cite{SM:BilbaoCrystallographicServer,SM:Aroyo2006BCSII,SM:Elcoro2021MagTQC,SM:Xu2020MagTopological}.  Repeated ordered-state symbols in that table denote distinct parent-group embeddings or corepresentation channels.

\input{tables/tr_mpg_character_longtable.tex}

\section{Microscopic derivation of nonlinear Zeeman free energies}

This section first derives the field-dependent free energy of a compensated magnetic domain from a general constrained-spin Hamiltonian.  We then apply the same construction to the collinear two-sublattice order of MnF$_2$ and to the noncollinear four-sublattice order of MnTe$_2$.

\subsection{General constrained-spin expansion and symmetry constraints}

Let $i$ and $j$ label magnetic sites throughout the crystal, and let $\langle ij\rangle$ count every interacting unordered pair once; no restriction to one primitive cell or to a single neighbor shell is implied.  A general fixed-length spin Hamiltonian including SOC-induced anisotropic interactions is
\begin{equation}
\begin{aligned}
H_{\rm spin}[\{\bm e_i\};\bB]
&=H_0[\{\bm e_i\}]
-\mu_B\sum_i m_i\bB\cdot\bm e_i,\\
H_0[\{\bm e_i\}]
&=\sum_i\bm e_i^{\mathsf T}\boldsymbol A_i\bm e_i
+\sum_{\langle ij\rangle}\bm e_i^{\mathsf T}\boldsymbol C_{ij}\bm e_j
+H_0^{(\geq4)}.
\end{aligned}
\label{eq:general_spin_model}
\end{equation}
The vector $\bm e_i$ is a unit spin direction, $m_i$ is the dimensionless local-moment magnitude expressed in units of $\mu_B$, and $\boldsymbol A_i$ is symmetric because its antisymmetric part does not contribute to the energy.  The term $H_0^{(\geq4)}$ contains symmetry-allowed four-spin and higher even-order interactions.  The real pair matrix is decomposed as
\begin{equation}
\begin{aligned}
\boldsymbol C_{ij}&=J_{ij}\boldsymbol I+\boldsymbol\Gamma_{ij}+\mathcal D(\bm D_{ij}),\\
\mathcal D(\bm D)&=
\begin{pmatrix}
0&D_z&-D_y\\
-D_z&0&D_x\\
D_y&-D_x&0
\end{pmatrix},
\end{aligned}
\label{eq:exchange_decomposition_supp}
\end{equation}
so that $\bm e_i^{\mathsf T}\mathcal D(\bm D_{ij})\bm e_j=\bm D_{ij}\cdot(\bm e_i\times\bm e_j)$.  Thus $J_{ij}$ is isotropic exchange, $\bm D_{ij}$ is the Dzyaloshinskii--Moriya vector, and $\boldsymbol\Gamma_{ij}$ is symmetric traceless anisotropic exchange.  The antisymmetric interaction follows the standard constructions of Refs.~\cite{SM:Dzyaloshinsky1958,SM:Moriya1960}.

The microscopic coefficients are constrained by the crystallographic symmetry.  Let $g=\{\boldsymbol R_g|\bm t_g\}$ map site $i$ to $g(i)$, and define the axial-vector matrix
\begin{equation}
\boldsymbol S_g=\det(\boldsymbol R_g)\boldsymbol R_g.
\label{eq:axial_spin_matrix_supp}
\end{equation}
A unitary operation sends $\bm e_i\mapsto \boldsymbol S_g\bm e_i$; an antiunitary operation adds a minus sign, which cancels from every even-order zero-field interaction.  Invariance of the bilinear terms requires
\begin{equation}
\begin{aligned}
m_{g(i)}&=m_i,\\
\boldsymbol A_{g(i)}&=\boldsymbol S_g\boldsymbol A_i\boldsymbol S_g^{\mathsf T},\\
\boldsymbol C_{g(i)g(j)}&=\boldsymbol S_g\boldsymbol C_{ij}\boldsymbol S_g^{\mathsf T},
\qquad \boldsymbol C_{ji}=\boldsymbol C_{ij}^{\mathsf T}.
\end{aligned}
\label{eq:general_pair_covariance_supp}
\end{equation}
More generally, if an $r$-spin interaction is written as $\mathcal J^{(r)}_{i_1\cdots i_r;\alpha_1\cdots\alpha_r}e_{i_1\alpha_1}\cdots e_{i_r\alpha_r}$, then
\begin{equation}
\mathcal J^{(r)}_{g(i_1)\cdots g(i_r);\alpha_1\cdots\alpha_r}
=\prod_{s=1}^{r}(\boldsymbol S_g)_{\alpha_s\beta_s}
\mathcal J^{(r)}_{i_1\cdots i_r;\beta_1\cdots\beta_r}.
\label{eq:general_multispin_covariance_supp}
\end{equation}
Equations~\eqref{eq:general_pair_covariance_supp} and \eqref{eq:general_multispin_covariance_supp} are the general symmetry constraints used in the two material reductions below.

\subsubsection{Constrained expansion about a compensated equilibrium pair.}

Let the two compensated zero-field domains be
\begin{equation}
\bm e_i^0(\eta)=\eta\bm n_i,
\qquad \eta=\pm1,
\qquad \sum_i m_i\bm n_i=0.
\label{eq:aiao_reference_supp}
\end{equation}
For a field $\bB$, the scalar branch free energy is the local minimum connected continuously to the chosen zero-field domain,
\begin{equation}
F_\eta(\bB)=
\min_{\{\bm e_i\}\in\mathcal N_\eta}
H_{\rm spin}[\{\bm e_i\};\bB],
\label{eq:branch_free_energy_supp}
\end{equation}
where $\mathcal N_\eta$ denotes that domain basin on the product of unit spheres.  This definition is the precise bridge from the lattice spin Hamiltonian to the scalar free-energy branch used in the Landau expansion.
We use the static-response convention
\begin{equation}
M_{\eta,i}=-\frac{\partial F_\eta}{\partial B_i}
=\chi_{\eta,ij}^{(1)}B_j
+\frac{1}{2!}\chi_{\eta,ijk}^{(2)}B_jB_k+\cdots,
\qquad
\chi_{\eta,ijk}^{(2)}
=-\left.\frac{\partial^3F_\eta}
{\partial B_i\partial B_j\partial B_k}\right|_{\bB=0}.
\label{eq:nonlinear_susceptibility_convention_supp}
\end{equation}
The factor $1/2!$ is therefore part of our definition of $\chi^{(2)}$.  Because it is obtained from derivatives of an equilibrium scalar free energy, $\chi_{\eta,ijk}^{(2)}$ is symmetric under permutations of its three indices.
We use $\delta\bm e_i$ for the tangent displacement throughout; no second symbol for the same quantity is introduced.  The conditions
\begin{equation}
\bm e_i^0\cdot\delta\bm e_i=0,
\qquad
\bm e_i=\bm e_i^0\sqrt{1-|\delta\bm e_i|^2}+\delta\bm e_i
\label{eq:normalized_tangent_map_supp}
\end{equation}
enforce $|\bm e_i|=1$ exactly.  The complete free energy in these coordinates is
\begin{equation}
\begin{aligned}
F_\eta[\{\delta\bm e_i\};\bB]={}&
H_0\!\left[\left\{\bm e_i^0\sqrt{1-|\delta\bm e_i|^2}
+\delta\bm e_i\right\}\right]\\
&-\mu_B\sum_i m_i\bB\cdot
\left(\bm e_i^0\sqrt{1-|\delta\bm e_i|^2}
+\delta\bm e_i\right).
\end{aligned}
\label{eq:complete_tangent_free_energy_supp}
\end{equation}
This expression contains the zero-field spin Hamiltonian, the Zeeman term, and the unit-length constraint before any expansion is made.

At zero field, direct expansion of the first line of Eq.~\eqref{eq:complete_tangent_free_energy_supp} defines the tangent stiffness and the cubic tangent form by
\begin{equation}
\begin{aligned}
H_0={}&H_0^{(0)}
+\frac12\sum_{ij}\delta\bm e_i^{\mathsf T}\boldsymbol K_{ij}\delta\bm e_j\\
&+\frac16\mathcal T_\eta[\delta\bm e,\delta\bm e,\delta\bm e]
+O(\delta e^4).
\end{aligned}
\label{eq:zero_field_tangent_expansion_supp}
\end{equation}
There is no linear term because $\{\bm e_i^0\}$ is a constrained zero-field equilibrium.  Equation~\eqref{eq:zero_field_tangent_expansion_supp} is a definition on the product of tangent planes: $\boldsymbol K_{ij}$ and $\mathcal T_\eta$ are determined by $H_0$ and the reference spin configuration and are not additional material parameters.

For comparison with a Lagrange-multiplier derivation, consider the same field-dependent problem in the form
\begin{equation}
\mathcal L=H_0-\mu_B\sum_i m_i\bB\cdot\bm e_i
+\frac12\sum_i\xi_i(\bm e_i^{\mathsf T}\bm e_i-1).
\label{eq:lagrange_function_supp}
\end{equation}
At either zero-field equilibrium,
\begin{equation}
\left.\frac{\partial H_0}{\partial\bm e_i}\right|_0
+\xi_i^0\bm e_i^0=0,
\qquad
\xi_i^0=-\bm e_i^0\cdot
\left.\frac{\partial H_0}{\partial\bm e_i}\right|_0.
\label{eq:zero_field_multiplier_supp}
\end{equation}
Applying the tangent projector to the first equality removes the multiplier.  Conversely, if the projected gradient vanishes, the gradient is parallel to $\bm e_i^0$, and choosing the second expression for $\xi_i^0$ recovers the full vector equation.  The projected and Lagrange-multiplier stationarity conditions are therefore equivalent.

Define
\begin{equation}
\boldsymbol\Pi_i=\boldsymbol I-\bm n_i\bm n_i^{\mathsf T}.
\label{eq:tangent_projector_supp}
\end{equation}
$\bm n_i\bm n_i^{\mathsf T}$ is a rank-one projector, not the identity: $\boldsymbol\Pi_i\bm n_i=0$, whereas $\boldsymbol\Pi_i\bm v=\bm v$ for every $\bm v\perp\bm n_i$.  Thus $\boldsymbol\Pi_i$ retains precisely the two physical rotation directions of spin $i$.  The constrained Hessian blocks defined in Eq.~\eqref{eq:zero_field_tangent_expansion_supp} can equivalently be evaluated from Eq.~\eqref{eq:lagrange_function_supp} as
\begin{equation}
\boldsymbol K_{ij}=\boldsymbol\Pi_i\left[
\left.\frac{\partial^2H_0}{\partial\bm e_i\partial\bm e_j}\right|_0
+\xi_i^0\delta_{ij}\boldsymbol I\right]\boldsymbol\Pi_j.
\label{eq:constrained_hessian_supp}
\end{equation}
The term proportional to $\xi_i^0$ is generated automatically in the direct expansion by the longitudinal correction $-|\delta\bm e_i|^2\bm e_i^0/2$.  The direct tangent expansion and the projected Cartesian Hessian therefore give the same $\boldsymbol K_{ij}$; no independent multiplier or stiffness parameter is introduced.

The cubic form in Eq.~\eqref{eq:zero_field_tangent_expansion_supp} is evaluated by the third tangent derivative
\begin{equation}
\mathcal T_\eta[\delta\bm e,\delta\bm e,\delta\bm e]
=\left.\frac{d^3}{dt^3}
H_0\!\left[\left\{\bm e_i^0\sqrt{1-t^2|\delta\bm e_i|^2}
+t\delta\bm e_i\right\}\right]\right|_{t=0}.
\label{eq:third_tangent_derivative_supp}
\end{equation}
The covariance conditions in Eqs.~\eqref{eq:general_pair_covariance_supp} and \eqref{eq:general_multispin_covariance_supp} induce the corresponding covariance of $\boldsymbol K$ and $\mathcal T_\eta$ in the tangent space.  Time-reversal invariance gives $\boldsymbol K^{(-\eta)}=\boldsymbol K^{(\eta)}$ and $\mathcal T_{-\eta}=-\mathcal T_\eta$ after the two domain tangent spaces are identified.

At finite field the constrained stationarity condition is
\begin{equation}
\left(\boldsymbol I-\bm e_i\bm e_i^{\mathsf T}\right)
\left(\frac{\partial H_0}{\partial\bm e_i}-\mu_Bm_i\bB\right)=0.
\label{eq:projected_stationarity_supp}
\end{equation}
Expanding the complete free energy about $\bm e_i^0$ through quadratic order gives
\begin{equation}
\begin{aligned}
F_\eta={}&F_\eta^{(0)}
+\frac12\sum_{ij}\delta\bm e_i^{\mathsf T}\boldsymbol K_{ij}\delta\bm e_j\\
&-\mu_B\sum_i m_i(\boldsymbol\Pi_i\bB)\cdot\delta\bm e_i+\cdots .
\end{aligned}
\label{eq:quadratic_tangent_free_energy_supp}
\end{equation}
Differentiating Eq.~\eqref{eq:quadratic_tangent_free_energy_supp}, or equivalently linearizing Eq.~\eqref{eq:projected_stationarity_supp}, gives
\begin{equation}
\sum_j\boldsymbol K_{ij}\delta\bm e_j^{(1)}=\mu_Bm_i\boldsymbol\Pi_i\bB,
\qquad
\delta\bm e_i^{(1)}=\mu_B\sum_j(\boldsymbol K^{-1})_{ij}m_j\boldsymbol\Pi_j\bB.
\label{eq:linear_tangent_response_supp}
\end{equation}
Thus $\delta\bm e_i^{(1)}\sim B$ and $|\delta\bm e_i^{(1)}|^2\sim B^2$.  The inverse is taken in the full tangent space.  The corresponding relaxed quadratic energy is
\begin{equation}
F_\eta^{(2)}=-\frac{\mu_B^2}{2}\sum_{ij}m_im_j
(\boldsymbol\Pi_i\bB)^{\mathsf T}(\boldsymbol K^{-1})_{ij}(\boldsymbol\Pi_j\bB).
\label{eq:relaxed_quadratic_energy_supp}
\end{equation}
Only the tangent component $\boldsymbol\Pi_i\bB$ produces a first-order torque.  The off-diagonal blocks of $\boldsymbol K^{-1}$ transmit that torque between different magnetic sites, so the response of one spin generally depends on the field projections at all sites.  The inverse in Eqs.~\eqref{eq:linear_tangent_response_supp} and \eqref{eq:relaxed_quadratic_energy_supp} is therefore the inverse on the full $2N$-dimensional tangent space, not a collection of independent single-site inverses.  A locally stable reference state requires this constrained stiffness to be positive definite.  Because $\boldsymbol\Pi_i$, $\boldsymbol K_{ij}$, and $\xi_i^0$ are unchanged by $\eta\to-\eta$, Eq.~\eqref{eq:relaxed_quadratic_energy_supp} is domain even and cannot select between the two domains.

To obtain the next order, expand
\begin{equation}
\bm e_i=\bm e_i^0+\delta\bm e_i
-\frac12|\delta\bm e_i|^2\bm e_i^0+\cdots .
\label{eq:tangent_map_expansion_supp}
\end{equation}
For a general $H_0$, the complete cubic contribution evaluated at the linear response is
\begin{equation}
\begin{aligned}
F_\eta^{(3)}={}&\frac16
\mathcal T_\eta[\delta\bm e^{(1)},\delta\bm e^{(1)},\delta\bm e^{(1)}]\\
&+\frac{\mu_B}{2}\sum_i m_i(\bB\cdot\bm e_i^0)
|\delta\bm e_i^{(1)}|^2.
\end{aligned}
\label{eq:general_cubic_energy_supp}
\end{equation}
The first line contains both the geometric contribution generated by fixed-length normalization and any contribution from $H_0^{(\geq4)}$; the second line is the corresponding Zeeman term.  To see why no second-order canting is needed, write $\delta\bm e_i=\delta\bm e_i^{(1)}+\delta\bm e_i^{(2)}+\cdots$.  The cross term from the quadratic stiffness and the term linear in $\delta\bm e_i^{(2)}$ from the Zeeman energy combine into
\begin{equation}
\sum_i(\delta\bm e_i^{(2)})^{\mathsf T}
\left(\sum_j\boldsymbol K_{ij}\delta\bm e_j^{(1)}-\mu_Bm_i\boldsymbol\Pi_i\bB\right)=0
\label{eq:second_order_cancellation_supp}
\end{equation}
which vanishes identically by Eq.~\eqref{eq:linear_tangent_response_supp}.  All remaining terms containing $\delta\bm e_i^{(2)}$ are fourth order or higher.  Hence the relaxed cubic energy requires only the linear canting.  Since $\bm e_i^0$ and $\mathcal T_\eta$ change sign between the two time-reversed domains whereas $\delta\bm e_i^{(1)}$ does not, Eq.~\eqref{eq:general_cubic_energy_supp} is domain odd and cubic in $\bB$.

For the quadratic part $H_0^{(2)}$ of Eq.~\eqref{eq:general_spin_model}, define the unconstrained Cartesian Hessian
\begin{equation}
\boldsymbol H_{ij}=\left.
\frac{\partial^2H_0^{(2)}}{\partial\bm e_i\partial\bm e_j}
\right|_0.
\label{eq:cartesian_hessian_supp}
\end{equation}
Because $H_0^{(2)}$ is quadratic, its third Cartesian derivative vanishes.  Its cubic tangent contribution arises entirely from inserting the longitudinal normalization correction in Eq.~\eqref{eq:tangent_map_expansion_supp}.  The result can first be written in the compact form used in the earlier main-text derivation,
\begin{equation}
\begin{aligned}
F_{\eta,\mathrm{quad}}^{(3)}={}&
-\frac12\sum_{ij}|\delta\bm e_j^{(1)}|^2
(\delta\bm e_i^{(1)})^{\mathsf T}
\boldsymbol H_{ij}\bm e_j^0\\
&+\frac{\mu_B}{2}\sum_i m_i(\bB\cdot\bm e_i^0)
|\delta\bm e_i^{(1)}|^2.
\end{aligned}
\label{eq:compact_quadratic_cubic_energy_supp}
\end{equation}
For the pair convention of Eq.~\eqref{eq:general_spin_model}, $\boldsymbol H_{ii}=2\boldsymbol A_i$, while an unordered pair contributes $\boldsymbol H_{ij}=\boldsymbol C_{ij}$ and $\boldsymbol H_{ji}=\boldsymbol C_{ij}^{\mathsf T}$.  Consequently, the zero-field part of Eq.~\eqref{eq:compact_quadratic_cubic_energy_supp}, or equivalently the third tangent derivative in Eq.~\eqref{eq:general_cubic_energy_supp}, expands explicitly as
\begin{equation}
\begin{aligned}
\frac16\mathcal T_\eta[\delta\bm e^{(1)}]^3={}&
-\sum_i|\delta\bm e_i^{(1)}|^2
(\delta\bm e_i^{(1)})^{\mathsf T}\boldsymbol A_i\bm e_i^0\\
&-\frac12\sum_{\langle ij\rangle}|\delta\bm e_j^{(1)}|^2
(\delta\bm e_i^{(1)})^{\mathsf T}\boldsymbol C_{ij}\bm e_j^0\\
&-\frac12\sum_{\langle ij\rangle}|\delta\bm e_i^{(1)}|^2
(\bm e_i^0)^{\mathsf T}\boldsymbol C_{ij}\delta\bm e_j^{(1)}.
\end{aligned}
\label{eq:quadratic_model_cubic_geometry_supp}
\end{equation}

It remains to connect this microscopic expansion to the scalar field polynomial used in the symmetry classification.  Define the field-response vectors, which are completely fixed by $\boldsymbol K$,
\begin{equation}
\bm{\mathcal R}_{ia}=\mu_B\sum_j(\boldsymbol K^{-1})_{ij}m_j\boldsymbol\Pi_j\hat{\bm e}_a,
\qquad
\delta\bm e_i^{(1)}=\sum_{a=x,y,z}\bm{\mathcal R}_{ia}B_a.
\label{eq:general_response_vectors_supp}
\end{equation}
Here $\hat{\bm e}_a$ is a Cartesian unit vector; $\bm{\mathcal R}_{ia}$ is a response derived from the microscopic Hamiltonian, not a new fitted coefficient.  Substitution into Eq.~\eqref{eq:general_cubic_energy_supp} gives
\begin{equation}
F_\eta^{(3)}=\sum_{abc}\Lambda_{abc}^{(\eta)}B_aB_bB_c,
\label{eq:general_cubic_response_tensor_supp}
\end{equation}
where the fully symmetric cubic-response tensor is
\begin{equation}
\begin{aligned}
\Lambda_{abc}^{(\eta)}={}&
\frac16\mathcal T_\eta[\bm{\mathcal R}_a,
\bm{\mathcal R}_b,\bm{\mathcal R}_c]\\
&+\frac{\mu_B}{6}\sum_i m_i\Bigl[
e_{ia}^0\,\bm{\mathcal R}_{ib}\cdot\bm{\mathcal R}_{ic}
+e_{ib}^0\,\bm{\mathcal R}_{ic}\cdot\bm{\mathcal R}_{ia}\\
&\hspace{42mm}
+e_{ic}^0\,\bm{\mathcal R}_{ia}\cdot\bm{\mathcal R}_{ib}
\Bigr].
\end{aligned}
\label{eq:general_cubic_tensor_coefficient_supp}
\end{equation}
In the first line, $\bm{\mathcal R}_a$ denotes the collection of all site vectors $\{\bm{\mathcal R}_{ia}\}$.  Time reversal gives $\Lambda_{abc}^{(-\eta)}=-\Lambda_{abc}^{(\eta)}$.  The parent magnetic point group then projects this tensor onto the cubic field covariants that transform like the one-dimensional domain character.  If $\{\phi_\alpha^{3}(\bB)\}$ is a basis of those covariants, the general result takes the scalar form
\begin{equation}
F_\eta^{(3)}=-\eta\sum_\alpha
\lambda_\alpha \phi_\alpha^{3}(\bB).
\label{eq:general_microscopic_selector_supp}
\end{equation}
Equations~\eqref{eq:general_response_vectors_supp}--\eqref{eq:general_microscopic_selector_supp} separate the two roles cleanly: magnetic symmetry fixes the allowed polynomials $\phi_\alpha^{3}$, while the microscopic Hamiltonian fixes their coefficients $\lambda_\alpha$.  The two material reductions below show how the same general expansion becomes a three-parameter transverse response for collinear MnF$_2$ and a seven-parameter response for noncollinear MnTe$_2$.

The crystal structures shown in the main-text figures were rendered with VESTA~\cite{SM:MommaIzumi2011VESTA}.

\subsection{Collinear \texorpdfstring{MnF$_2$}{MnF2}: two-sublattice reduction}

\subsubsection{From the lattice Hamiltonian to the uniform model.}

Rutile MnF$_2$ has nonmagnetic space group $P4_2/mnm$ and an easy-axis antiferromagnetic state with the moments along the tetragonal axis~\cite{SM:AlteDaVeiga1982MnF2,SM:Higuchi2016MnF2,SM:Bhowal2024Octupoles}.  Its ordered-state MPG is $\mathcal M=4'/mm'm$, and adjoining pure time reversal gives the gray parent $\mathcal G=4/mmm1'$.  The two magnetic sublattices are denoted by $a$ and $b$, and $\bm S_{\mu\bm R}$ is the unit spin direction on sublattice $\mu=a,b$ in cell $\bm R$.  Before imposing uniformity, the lattice Hamiltonian can be written as
\begin{equation}
\begin{aligned}
\mathcal H_0={}&
\sum_{\bm R,\mu}
\bm S_{\mu\bm R}^{\mathsf T}
\boldsymbol A_{\kappa,\mu}^{(0)}
\bm S_{\mu\bm R}\\
&+\frac12
\sum_{\bm R,\bm R'}\sum_{\mu,\nu}
\bm S_{\mu\bm R}^{\mathsf T}
\boldsymbol A_{J,\mu\nu}(\bm R-\bm R')
\bm S_{\nu\bm R'} .
\end{aligned}
\label{eq:mnf2_lattice_model_supp}
\end{equation}
The matrices $\boldsymbol A_{\kappa,\mu}^{(0)}$ and $\boldsymbol A_{J,\mu\nu}$ are effective spin interactions; microscopic hopping and spin--orbit effects are encoded in their matrix elements.  To describe the spatially uniform response, we assume that the spin direction on a given sublattice is identical in every unit cell, $\bm S_{\mu\bm R}=\bm S_\mu$.  This uniform-sublattice approximation is the only reduction made in passing to the main-text model: no neighbor shell is truncated, because all real-space bonds are retained in the cell-averaged tensors.  Writing $\bm d=\bm R-\bm R'$ and using $\boldsymbol A_{J,\nu\mu}(-\bm d)=\boldsymbol A_{J,\mu\nu}(\bm d)^{\mathsf T}$, define
\begin{equation}
\begin{aligned}
\boldsymbol A_\kappa^\mu&=
\boldsymbol A_{\kappa,\mu}^{(0)}
+\frac14\sum_{\bm d}
\left[\boldsymbol A_{J,\mu\mu}(\bm d)
+\boldsymbol A_{J,\mu\mu}(\bm d)^{\mathsf T}\right],\\
\boldsymbol A_J&=\sum_{\bm d}\boldsymbol A_{J,ab}(\bm d).
\end{aligned}
\label{eq:mnf2_uniform_tensor_sums_supp}
\end{equation}
The first line absorbs every same-sublattice bond into a symmetric effective on-site tensor, while the second line sums all intersublattice bonds.  Dividing by the number of primitive cells then gives
\begin{equation}
U_0=
\bm S_a^{\mathsf T}\boldsymbol A_\kappa^a\bm S_a
+\bm S_b^{\mathsf T}\boldsymbol A_\kappa^b\bm S_b
+\bm S_a^{\mathsf T}\boldsymbol A_J\bm S_b .
\label{eq:mnf2_uniform_general_supp}
\end{equation}

Only the symmetric part of an on-site tensor contributes to a quadratic form, whereas the intersublattice tensor need not initially be symmetric.  Before imposing the MPG constraints, we therefore write
\begin{equation}
\boldsymbol A_\kappa^a=
\begin{pmatrix}
a_{xx}&a_{xy}&a_{xz}\\
a_{xy}&a_{yy}&a_{yz}\\
a_{xz}&a_{yz}&a_{zz}
\end{pmatrix},
\quad
\boldsymbol A_\kappa^b=
\begin{pmatrix}
b_{xx}&b_{xy}&b_{xz}\\
b_{xy}&b_{yy}&b_{yz}\\
b_{xz}&b_{yz}&b_{zz}
\end{pmatrix},
\quad
\boldsymbol A_J=
\begin{pmatrix}
j_{xx}&j_{xy}&j_{xz}\\
j_{yx}&j_{yy}&j_{yz}\\
j_{zx}&j_{zy}&j_{zz}
\end{pmatrix}.
\label{eq:mnf2_unconstrained_tensors_supp}
\end{equation}
Thus the uniform model starts from 21 real coefficients: six in each on-site tensor and nine in $\boldsymbol A_J$.

A convenient set of operations for reducing these tensors is $C_{4z}\mathcal T$, $C_{2z}$, and $C_{2x}\mathcal T$; inversion acts trivially on an axial spin and adds no independent uniform-spin constraint.  On spin components, the relevant matrices are
\begin{equation}
\boldsymbol C_{4z}=
\begin{pmatrix}
0&-1&0\\
1&0&0\\
0&0&1
\end{pmatrix},
\qquad
\boldsymbol Q_z=\boldsymbol C_{2z}=\operatorname{diag}(-1,-1,1),
\qquad
\boldsymbol Q_x=-\boldsymbol C_{2x}=\operatorname{diag}(-1,1,1).
\label{eq:mnf2_generator_matrices_supp}
\end{equation}
The minus sign in $\boldsymbol Q_x$ is the spin reversal supplied by $\mathcal T$.  The operation $C_{4z}\mathcal T$ exchanges $a$ and $b$, whereas $C_{2z}$ and $C_{2x}\mathcal T$ leave each sublattice invariant up to a lattice translation.  Since every term is bilinear in spins, the two time-reversal signs cancel.  Invariance therefore requires
\begin{equation}
\begin{aligned}
\boldsymbol A_\kappa^b&=\boldsymbol C_{4z}^{\mathsf T}\boldsymbol A_\kappa^a\boldsymbol C_{4z},
&\boldsymbol A_J&=\boldsymbol C_{4z}^{\mathsf T}\boldsymbol A_J^{\mathsf T}\boldsymbol C_{4z},\\
\boldsymbol A_\kappa^\mu&=\boldsymbol Q_z^{\mathsf T}\boldsymbol A_\kappa^\mu\boldsymbol Q_z,
&\boldsymbol A_J&=\boldsymbol Q_z^{\mathsf T}\boldsymbol A_J\boldsymbol Q_z,\\
\boldsymbol A_\kappa^\mu&=\boldsymbol Q_x^{\mathsf T}\boldsymbol A_\kappa^\mu\boldsymbol Q_x,
&\boldsymbol A_J&=\boldsymbol Q_x^{\mathsf T}\boldsymbol A_J\boldsymbol Q_x,
\qquad \mu=a,b.
\end{aligned}
\label{eq:mnf2_tensor_constraints_supp}
\end{equation}
The $\boldsymbol Q_z$ equations first set
\begin{equation}
a_{xz}=a_{yz}=b_{xz}=b_{yz}=0,
\qquad
j_{xz}=j_{zx}=j_{yz}=j_{zy}=0,
\end{equation}
and the $Q_x$ equations then set
\begin{equation}
a_{xy}=b_{xy}=j_{xy}=j_{yx}=0.
\end{equation}
The $C_{4z}\mathcal T$ equations finally exchange the two transverse on-site coefficients and equate the two transverse exchange coefficients.  The complete symmetry-reduced tensors are therefore
\begin{equation}
\boxed{
\boldsymbol A_\kappa^a=\operatorname{diag}(\kappa_x,\kappa_y,\kappa_z),
\qquad
\boldsymbol A_\kappa^b=\operatorname{diag}(\kappa_y,\kappa_x,\kappa_z),
\qquad
\boldsymbol A_J=\operatorname{diag}(J_\perp,J_\perp,J_z).}
\label{eq:mnf2_reduced_tensors_supp}
\end{equation}
Substitution into Eq.~\eqref{eq:mnf2_uniform_general_supp} gives
\begin{equation}
\begin{aligned}
U_0={}&
\kappa
\left(S_{a,x}^2+S_{a,y}^2+S_{b,x}^2+S_{b,y}^2\right)\\
&+\delta\left[
(S_{a,x}^2-S_{a,y}^2)
-(S_{b,x}^2-S_{b,y}^2)
\right]\\
&+\kappa_z(S_{a,z}^2+S_{b,z}^2)\\
&+J_\perp(S_{a,x}S_{b,x}+S_{a,y}S_{b,y})
+J_zS_{a,z}S_{b,z},
\end{aligned}
\label{eq:mnf2_reduced_model_supp}
\end{equation}
where
\begin{equation}
\kappa=\frac{\kappa_x+\kappa_y}{2},
\qquad
\delta=\frac{\kappa_x-\kappa_y}{2}.
\label{eq:mnf2_delta_definition_supp}
\end{equation}
The symmetry $C_{4z}\mathcal T$ relates the two transverse channels but does not require them to be equal, so it permits $\delta\ne0$.  If $\mathcal{PT}$ is restored, it maps $\bm S_a\mapsto-\bm S_b$ and $\bm S_b\mapsto-\bm S_a$.  Equation~\eqref{eq:mnf2_uniform_general_supp} must then also satisfy
\begin{equation}
\boldsymbol A_\kappa^a=\boldsymbol A_\kappa^b,
\qquad
\boldsymbol A_J=\boldsymbol A_J^{\mathsf T}.
\end{equation}
Together with Eq.~\eqref{eq:mnf2_reduced_tensors_supp}, this gives $\kappa_x=\kappa_y$ and hence $\delta=0$.  Independently, the MPG covariance equation selects the cubic polynomial
\begin{equation}
\phi^3(\bB)=B_z(B_x^2-B_y^2).
\label{eq:mnf2_selector_supp}
\end{equation}

\subsubsection{Field-induced stationary moments.}

The two zero-field domains are
\begin{equation}
\bm S_a^{(0)}=\eta\hat{\bm z},
\qquad
\bm S_b^{(0)}=-\eta\hat{\bm z},
\qquad
\eta=\pm1.
\label{eq:mnf2_domains_supp}
\end{equation}
Let $h$ be the magnetic-moment magnitude of each Mn site, so that the actual moments are $h\bm S_a$ and $h\bm S_b$.  In a magnetic field,
\begin{equation}
F=U_0-h\bB\cdot(\bm S_a+\bm S_b).
\label{eq:mnf2_field_energy_supp}
\end{equation}
The fixed-length constraints are imposed exactly by
\begin{equation}
\begin{aligned}
\bm S_a&=
\left(S_{a,x},S_{a,y},
\eta\sqrt{1-S_{a,x}^2-S_{a,y}^2}\right),\\
\bm S_b&=
\left(S_{b,x},S_{b,y},
-\eta\sqrt{1-S_{b,x}^2-S_{b,y}^2}\right).
\end{aligned}
\label{eq:mnf2_exact_parameterization_supp}
\end{equation}
Expanding Eq.~\eqref{eq:mnf2_field_energy_supp} to the order required for the cubic field response gives
\begin{equation}
\begin{aligned}
F={}&F_0
+(K+\delta)(S_{a,x}^2+S_{b,y}^2)
+(K-\delta)(S_{a,y}^2+S_{b,x}^2)\\
&+J_\perp(S_{a,x}S_{b,x}+S_{a,y}S_{b,y})\\
&-hB_x(S_{a,x}+S_{b,x})
-hB_y(S_{a,y}+S_{b,y})\\
&+\frac{\eta hB_z}{2}
\left(S_{a,x}^2+S_{a,y}^2-S_{b,x}^2-S_{b,y}^2\right)
+O(B^4),
\end{aligned}
\label{eq:mnf2_expanded_field_energy_supp}
\end{equation}
where
\begin{equation}
K=\kappa-\kappa_z+\frac{J_z}{2}.
\label{eq:mnf2_K_definition_supp}
\end{equation}
The transverse components are $O(B)$.  Consequently, the last line of Eq.~\eqref{eq:mnf2_expanded_field_energy_supp} is cubic in the field and does not alter their first-order solution.

Variation with respect to the four transverse components gives
\begin{equation}
\begin{aligned}
2(K+\delta)S_{a,x}+J_\perp S_{b,x}&=hB_x,\\
J_\perp S_{a,x}+2(K-\delta)S_{b,x}&=hB_x,\\
2(K-\delta)S_{a,y}+J_\perp S_{b,y}&=hB_y,\\
J_\perp S_{a,y}+2(K+\delta)S_{b,y}&=hB_y.
\end{aligned}
\label{eq:mnf2_linear_systems_supp}
\end{equation}
Their solution is
\begin{equation}
\begin{aligned}
S_{a,x}'&=\alpha B_x,&
S_{a,y}'&=\beta B_y,\\
S_{b,x}'&=\beta B_x,&
S_{b,y}'&=\alpha B_y,
\end{aligned}
\label{eq:mnf2_transverse_solution_supp}
\end{equation}
with
\begin{equation}
\alpha=
\frac{h(2K-2\delta-J_\perp)}
{4K^2-4\delta^2-J_\perp^2},
\qquad
\beta=
\frac{h(2K+2\delta-J_\perp)}
{4K^2-4\delta^2-J_\perp^2}.
\label{eq:mnf2_alpha_beta_supp}
\end{equation}
Thus the stationary fixed-length moments are
\begin{equation}
\begin{aligned}
\bm S_a'={}&
\left(\alpha B_x,\beta B_y,
\eta\sqrt{1-\alpha^2B_x^2-\beta^2B_y^2}\right),\\
\bm S_b'={}&
\left(\beta B_x,\alpha B_y,
-\eta\sqrt{1-\beta^2B_x^2-\alpha^2B_y^2}\right).
\end{aligned}
\label{eq:mnf2_stationary_moments_supp}
\end{equation}
The operation $C_{4z}\mathcal T$ exchanges the two responses.  It permits $\alpha\ne\beta$ when $\delta\ne0$, but does not itself force this inequality.

\subsubsection{Longitudinal response and cubic free energy.}

Expanding the longitudinal components of Eq.~\eqref{eq:mnf2_stationary_moments_supp} gives
\begin{equation}
S_{a,z}'+S_{b,z}'
=\frac{\eta}{2}(\beta^2-\alpha^2)(B_x^2-B_y^2)
+O(B^4).
\label{eq:mnf2_longitudinal_sum_supp}
\end{equation}
The transverse tilts are linear in the field, whereas the fixed-length constraint produces the quadratic longitudinal response in Eq.~\eqref{eq:mnf2_longitudinal_sum_supp}.  If $\alpha=\beta$, the two longitudinal corrections cancel.  If $\alpha\ne\beta$, they leave a domain-odd net component with the angular form $B_x^2-B_y^2$.  Its ordinary Zeeman coupling to $B_z$ is therefore cubic.  A $B_z$-induced second-order correction to the transverse stationary point does not generate an additional cubic term: the first-order solution already makes the quadratic energy stationary, so inserting that correction changes the stationary free energy only from $O(B^4)$ onward.

Substituting Eq.~\eqref{eq:mnf2_stationary_moments_supp} directly into Eq.~\eqref{eq:mnf2_field_energy_supp} yields
\begin{equation}
\begin{aligned}
F_\eta(\bB)={}&F_0
-\frac{h^2(2K-J_\perp)}
{4K^2-4\delta^2-J_\perp^2}
(B_x^2+B_y^2)\\
&-\eta\lambda B_z(B_x^2-B_y^2)+O(B^4),
\end{aligned}
\label{eq:mnf2_relaxed_free_energy_supp}
\end{equation}
where
\begin{equation}
\boxed{
\lambda=
\frac{h}{2}(\beta^2-\alpha^2)
=\frac{4h^3\delta(2K-J_\perp)}
{(4K^2-4\delta^2-J_\perp^2)^2}.}
\label{eq:mnf2_lambda_supp}
\end{equation}
With the convention in Eq.~\eqref{eq:nonlinear_susceptibility_convention_supp},
\begin{equation}
\chi_{\eta,zxx}^{(2)}=2\eta\lambda,
\qquad
\chi_{\eta,zyy}^{(2)}=-2\eta\lambda,
\label{eq:mnf2_kappa2_supp}
\end{equation}
together with their index permutations.  Equations~\eqref{eq:mnf2_alpha_beta_supp}--\eqref{eq:mnf2_lambda_supp} display the complete microscopic chain
\begin{equation}
\delta\ne0
\Longrightarrow \alpha\ne\beta
\Longrightarrow
S_{a,z}'+S_{b,z}'\propto\eta(B_x^2-B_y^2)
\Longrightarrow
F_\eta^{(3)}=-\eta\lambda B_z(B_x^2-B_y^2).
\end{equation}
In the $\mathcal{PT}$-symmetric limit,
\begin{equation}
\delta=0\Longrightarrow\alpha=\beta\Longrightarrow\lambda=0.
\label{eq:mnf2_PT_limit_supp}
\end{equation}
The transverse stationary point is stable when
\begin{equation}
K+\delta>0,
\qquad
K-\delta>0,
\qquad
4K^2-4\delta^2-J_\perp^2>0.
\label{eq:mnf2_stability_supp}
\end{equation}
Defining
\begin{equation}
\delta_{\rm st}=\frac12\sqrt{4K^2-J_\perp^2},
\qquad
\lambda_{\rm st}=
\frac{2h^3(2K-J_\perp)}
{(4K^2-J_\perp^2)^{3/2}},
\end{equation}
Eq.~\eqref{eq:mnf2_lambda_supp} can also be written as
\begin{equation}
\frac{\lambda}{\lambda_{\rm st}}
=\frac{\delta/\delta_{\rm st}}
{\left[1-(\delta/\delta_{\rm st})^2\right]^2}.
\label{eq:mnf2_scaled_lambda_supp}
\end{equation}
Hence $\lambda$ is odd in the $\mathcal{PT}$-breaking parameter and is linear in $\delta$ near the symmetric limit.

\subsection{Noncollinear \texorpdfstring{MnTe$_2$}{MnTe2}: four-sublattice reduction}

\subsubsection{Symmetry reduction and parameter count.}

The experimentally established AIAO domains of MnTe$_2$~\cite{SM:Burlet1997MnTe2} have single-domain magnetic point group $m\bar{3}$ and gray parent group $m\bar{3}1'$.  In the BCS MPOINT and COREPRESENTATIONS PG conventions~\cite{SM:BilbaoCrystallographicServer,SM:Aroyo2006BCS,SM:Aroyo2006BCSII,SM:BorovikRomanov2013,SM:Elcoro2021MagTQC,SM:Xu2020MagTopological}, the domain coordinate belongs to $\mathrm mA_g$, the axial field transforms as the three-component $\mathrm mT_g$ object, and the lowest domain-matching scalar is $B_xB_yB_z$.  The selected corepresentations are listed in the main text.

We use the lattice-locked Mn-site order
\begin{equation}
\begin{aligned}
\bm r_1&=(0,0,0),&
\bm r_2&=(\tfrac12,0,\tfrac12),\\
\bm r_3&=(\tfrac12,\tfrac12,0),&
\bm r_4&=(0,\tfrac12,\tfrac12),
\end{aligned}
\label{eq:mnte2_mn_sites_supp}
\end{equation}
and the corresponding AIAO axes
\begin{equation}
\begin{aligned}
\bm n_1&=(1,1,1)/\sqrt3,&
\bm n_2&=(-1,-1,1)/\sqrt3,\\
\bm n_3&=(1,-1,-1)/\sqrt3,&
\bm n_4&=(-1,1,-1)/\sqrt3.
\end{aligned}
\label{eq:canonical_aiao_axes_supp}
\end{equation}
This site--moment association is fixed in the presence of SOC.  For MnTe$_2$ we set $H_0^{(\geq4)}=0$ and retain the symmetry-complete bilinear model.  In the uniform response sector, $\bm e_i=(x_i,y_i,z_i)$ denotes the common spin direction on sublattice $i$.  All real-space interactions connecting a given pair of sublattices are summed into the corresponding effective matrix $\boldsymbol C_{ij}$, while same-sublattice quadratic sums are absorbed into $\boldsymbol A_i$.  The six pairs below are therefore the six unordered sublattice pairs, not a restriction to one neighbor shell.  Before imposing symmetry, the effective quadratic energy contains
\begin{equation}
H_0^{(2)}=
\sum_i\sum_{\alpha\leq\beta}\boldsymbol A_i^{\alpha\beta}e_{i\alpha}e_{i\beta}
+\sum_{\langle ij\rangle}\sum_{\alpha\beta}\boldsymbol C_{ij}^{\alpha\beta}e_{i\alpha}e_{j\beta}.
\label{eq:raw_scalar_quadratic_supp}
\end{equation}
There are $4\times6=24$ symmetric on-site coefficients and $6\times9=54$ pair coefficients, for 78 scalar coefficients in total.

The proper part of $T_h$ is generated by
\begin{equation}
\begin{aligned}
C_{3[111]}:&(x,y,z)\mapsto(y,z,x),
&&(1,2,3,4)\mapsto(1,4,2,3),\\
C_{2z}:&(x,y,z)\mapsto(-x,-y,z),
&&(1,2,3,4)\mapsto(2,1,4,3).
\end{aligned}
\label{eq:th_generators_spin_supp}
\end{equation}
For axial spins in this uniform four-sublattice model, inversion duplicates the action of a proper operation and adds no independent constraint.  If $g$ permutes the sublattices and acts on an axial spin by $\boldsymbol R_g^{\rm ax}=\det(\boldsymbol R_g)\boldsymbol R_g$, invariance requires
\begin{equation}
\boldsymbol C_{g(i)g(j)}=\boldsymbol R_g^{\rm ax}\boldsymbol C_{ij}(\boldsymbol R_g^{\rm ax})^{\mathsf T},
\qquad
\boldsymbol A_{g(i)}=\boldsymbol R_g^{\rm ax}\boldsymbol A_i(\boldsymbol R_g^{\rm ax})^{\mathsf T},
\label{eq:microscopic_covariance_supp}
\end{equation}
with $\boldsymbol C_{ji}=\boldsymbol C_{ij}^{\mathsf T}$.

We first apply Eq.~\eqref{eq:microscopic_covariance_supp} to the on-site terms.  Before symmetry reduction, write
\begin{equation}
\boldsymbol A_1^{\rm gen}=
\begin{pmatrix}
a_1&b_1&c_1\\
b_1&d_1&e_1\\
c_1&e_1&f_1
\end{pmatrix},
\qquad
\boldsymbol R_3=
\begin{pmatrix}
0&1&0\\
0&0&1\\
1&0&0
\end{pmatrix},
\label{eq:mnte2_general_onsite_supp}
\end{equation}
where $\boldsymbol R_3(x,y,z)^{\mathsf T}=(y,z,x)^{\mathsf T}$.  Site 1 is fixed by $C_{3[111]}$, so $\boldsymbol A_1=\boldsymbol R_3\boldsymbol A_1\boldsymbol R_3^{\mathsf T}$.  Comparing matrix elements gives $a_1=d_1=f_1\equiv a$ and $b_1=c_1=e_1\equiv b$, hence
\begin{equation}
\boldsymbol A_1=
\begin{pmatrix}
a&b&b\\
b&a&b\\
b&b&a
\end{pmatrix}.
\label{eq:onsite_a1_supp}
\end{equation}
The other three matrices follow by the sublattice permutations in Eq.~\eqref{eq:th_generators_spin_supp}; explicitly,
\begin{equation}
\begin{aligned}
\boldsymbol A_2&=
\begin{pmatrix}a&b&-b\\ b&a&-b\\ -b&-b&a\end{pmatrix},
&
\boldsymbol A_3&=
\begin{pmatrix}a&-b&-b\\ -b&a&b\\ -b&b&a\end{pmatrix},\\[4pt]
\boldsymbol A_4&=
\begin{pmatrix}a&-b&b\\ -b&a&-b\\ b&-b&a\end{pmatrix}.
\end{aligned}
\label{eq:mnte2_all_onsite_tensors_supp}
\end{equation}
Together with Eq.~\eqref{eq:onsite_a1_supp}, these are the complete four on-site tensors.  The $a$ term is proportional to $\sum_i|\bm e_i|^2=4$ and is therefore constant.  A convenient basis for the remaining on-site invariant is
\begin{equation}
H_A=-A\sum_i(\bm e_i\cdot\bm n_i)^2.
\label{eq:onsite_aiao_anisotropy_supp}
\end{equation}
Its scalar content is explicit from
\begin{equation}
\sum_i(\bm e_i\cdot\bm n_i)^2=\frac43+\frac23Q_A,
\label{eq:onsite_scalar_relation_supp}
\end{equation}
where
\begin{equation}
\begin{aligned}
Q_A={}&(x_1y_1+x_1z_1+y_1z_1)
+(x_2y_2-x_2z_2-y_2z_2)\\
&+(-x_3y_3-x_3z_3+y_3z_3)
+(-x_4y_4+x_4z_4-y_4z_4).
\end{aligned}
\label{eq:onsite_scalar_polynomial_supp}
\end{equation}

Next consider the representative pair $12$.  Because $C_{2z}$ exchanges its endpoints,
\begin{equation}
\boldsymbol C_{12}=\boldsymbol R_{2z}\boldsymbol C_{12}^{\mathsf T}\boldsymbol R_{2z}^{\mathsf T},
\qquad \boldsymbol R_{2z}=\operatorname{diag}(-1,-1,1).
\label{eq:bond12_constraint_supp}
\end{equation}
For a general $3\times3$ matrix this gives
\begin{equation}
\boldsymbol C_{12}^{xy}=\boldsymbol C_{12}^{yx},\qquad
\boldsymbol C_{12}^{xz}=-\boldsymbol C_{12}^{zx},\qquad
\boldsymbol C_{12}^{yz}=-\boldsymbol C_{12}^{zy}.
\label{eq:bond12_component_constraints_supp}
\end{equation}
The six surviving coefficients can be written as
\begin{equation}
\boldsymbol C_{12}=
\begin{pmatrix}
J+\gamma_x&\gamma_{xy}&-D_y\\
\gamma_{xy}&J+\gamma_y&D_x\\
D_y&-D_x&J-\gamma_x-\gamma_y
\end{pmatrix}.
\label{eq:bond12_matrix_supp}
\end{equation}
The decomposition used here is
\begin{equation}
\boldsymbol C_{ij}=J\boldsymbol I+\boldsymbol\Gamma_{ij}+\mathcal D(\bm D_{ij}),
\qquad
\mathcal D(\bm D)=
\begin{pmatrix}
0&D_z&-D_y\\
-D_z&0&D_x\\
D_y&-D_x&0
\end{pmatrix},
\label{eq:mnte2_exchange_decomposition_explicit_supp}
\end{equation}
where $\boldsymbol\Gamma_{ij}$ is symmetric and traceless, while $\mathcal D(\bm D_{ij})$ is the antisymmetric Dzyaloshinskii--Moriya part.
Equivalently, the scalar energy of this representative effective pair is
\begin{equation}
\begin{aligned}
H_{12}={}&(J+\gamma_x)x_1x_2+(J+\gamma_y)y_1y_2
+(J-\gamma_x-\gamma_y)z_1z_2\\
&+\gamma_{xy}(x_1y_2+y_1x_2)
+D_x(y_1z_2-z_1y_2)\\
&+D_y(z_1x_2-x_1z_2).
\end{aligned}
\label{eq:bond12_scalar_supp}
\end{equation}
Thus $D_z$, $\Gamma_{xz}$, and $\Gamma_{yz}$ vanish for this pair.  The two generators produce the other five pairs, listed in Table~\ref{tab:th_exchange_tensors_supp}.

\begin{table}[htbp]
\caption{Complete exchange matrices for the six directed representative pairs in the four-sublattice MnTe$_2$ model.  The reverse orientation is $\boldsymbol C_{ji}=\boldsymbol C_{ij}^{\mathsf T}$.}
\label{tab:th_exchange_tensors_supp}
\centering
\footnotesize
\renewcommand{\arraystretch}{1.6}
\begin{tabular}{c c}
\hline
pair $ij$ & full matrix $\boldsymbol C_{ij}$\\
\hline
$12$ & $\begin{pmatrix}J+\gamma_x&\gamma_{xy}&-D_y\\ \gamma_{xy}&J+\gamma_y&D_x\\ D_y&-D_x&J-\gamma_x-\gamma_y\end{pmatrix}$\\
$13$ & $\begin{pmatrix}J-\gamma_x-\gamma_y&D_y&-D_x\\ -D_y&J+\gamma_x&\gamma_{xy}\\ D_x&\gamma_{xy}&J+\gamma_y\end{pmatrix}$\\
$14$ & $\begin{pmatrix}J+\gamma_y&D_x&\gamma_{xy}\\ -D_x&J-\gamma_x-\gamma_y&D_y\\ \gamma_{xy}&-D_y&J+\gamma_x\end{pmatrix}$\\
$23$ & $\begin{pmatrix}J+\gamma_y&D_x&-\gamma_{xy}\\ -D_x&J-\gamma_x-\gamma_y&-D_y\\ -\gamma_{xy}&D_y&J+\gamma_x\end{pmatrix}$\\
$24$ & $\begin{pmatrix}J-\gamma_x-\gamma_y&D_y&D_x\\ -D_y&J+\gamma_x&-\gamma_{xy}\\ -D_x&-\gamma_{xy}&J+\gamma_y\end{pmatrix}$\\
$34$ & $\begin{pmatrix}J+\gamma_x&-\gamma_{xy}&D_y\\ -\gamma_{xy}&J+\gamma_y&D_x\\ -D_y&-D_x&J-\gamma_x-\gamma_y\end{pmatrix}$\\
\hline
\end{tabular}
\end{table}

The table lists one orientation for each pair.  Reversing an orientation gives
\begin{equation}
\bm D_{ji}=-\bm D_{ij},
\qquad
\boldsymbol\Gamma_{ji}=\boldsymbol\Gamma_{ij}^{\mathsf T},
\label{eq:directed_pair_reversal_supp}
\end{equation}
so, for example, $\bm D_{41}=-\bm D_{14}=(-D_y,0,-D_x)$.  No additional coefficient is introduced by a reversed pair.

The same reduction can be stated as a projection in the 78-dimensional coefficient space of Eq.~\eqref{eq:raw_scalar_quadratic_supp}.  Denote the 12-element proper tetrahedral subgroup by $T$.  If $U(g)$ is its induced action on those coefficients, the Reynolds projector
\begin{equation}
\boldsymbol\Pi_T=\frac1{12}\sum_{g\in T}\boldsymbol U(g)
\label{eq:reynolds_projector_spin_supp}
\end{equation}
has rank eight.  One projected direction is the constant $\sum_i|\bm e_i|^2$; removing it leaves seven physical quadratic invariants.  The resulting model is
\begin{equation}
H_0=
-A\sum_i(\bm e_i\cdot\bm n_i)^2
+\sum_{\langle ij\rangle}\bm e_i^{\mathsf T}
\left[J\boldsymbol I+\boldsymbol\Gamma_{ij}+\mathcal D(\bm D_{ij})\right]\bm e_j,
\label{eq:seven_parameter_th_model_supp}
\end{equation}
with the seven independent coefficients
\begin{equation}
\left(A,J,D_x,D_y,\gamma_x,\gamma_y,\gamma_{xy}\right).
\label{eq:seven_parameters_supp}
\end{equation}
\subsubsection{Canonical AIAO stationarity and the \texorpdfstring{$T_h$}{Th} two-mode reduction.}

Substituting the six pair tensors in Table~\ref{tab:th_exchange_tensors_supp} into the seven-parameter Hamiltonian gives, for every sublattice of the canonical AIAO state,
\begin{equation}
\left.\frac{\partial H_0}{\partial\bm e_i}\right|_0
=\eta\left(2D_x-2D_y-J-2A
-2\gamma_x-2\gamma_{xy}-2\gamma_y\right)\bm n_i.
\label{eq:aiao_cartesian_gradient_supp}
\end{equation}
The gradient is parallel to $\bm e_i^0$, confirming the constrained equilibrium used in the general expansion.  The corresponding multiplier is
\begin{equation}
\xi_i^0=J+2A-2D_x+2D_y
+2\gamma_x+2\gamma_{xy}+2\gamma_y,
\label{eq:explicit_zero_field_multiplier_supp}
\end{equation}
which is fixed by the seven coefficients and is not fitted independently.  Whether the equilibrium is a local minimum is determined separately by the full tangent Hessian.  With four symmetry-equivalent Mn moments, set $h=\mu_Bm$.

The two-mode form can be proved without assuming a trial canting pattern.  Write the most general linear response as $\delta\bm e_i^{(1)}=\boldsymbol L_i\bB$.  Tangency and covariance under a proper tetrahedral operation $g$ require
\begin{equation}
\bm n_i^{\mathsf T}\boldsymbol L_i=0,
\qquad
\boldsymbol L_{g(i)}\boldsymbol R_g=\boldsymbol R_g\boldsymbol L_i.
\label{eq:linear_map_equivariance_supp}
\end{equation}
The second relation is a covariance statement, not a symmetry of the configuration at a fixed generic field: it is equivalent to $\delta\bm e_{g(i)}^{(1)}(\boldsymbol R_g\bB)=\boldsymbol R_g\delta\bm e_i^{(1)}(\bB)$.  Thus $T_h$ constrains the derivative at $\bB=0$ even though a chosen nonzero $\bB$ generally lowers the symmetry.
Site 1 is fixed by $C_{3[111]}$.  Choose an oriented local basis $\{\bm n_1,\bm t_1,\bm t_2\}$ with $\bm t_2=\bm n_1\times\bm t_1$, and write
\begin{equation}
\bB=B_\parallel\bm n_1+B_1\bm t_1+B_2\bm t_2,
\qquad
\delta\bm e_1^{(1)}=x\bm t_1+y\bm t_2.
\label{eq:local_tangent_components_supp}
\end{equation}
Before imposing the site symmetry, the most general tangent linear response is
\begin{equation}
\begin{pmatrix}x\\y\end{pmatrix}
=\bm v B_\parallel+\boldsymbol M\begin{pmatrix}B_1\\B_2\end{pmatrix},
\label{eq:general_local_response_supp}
\end{equation}
where $\bm v$ is a real two-component vector and $\boldsymbol M$ is a real $2\times2$ matrix.  The $C_3$ rotation acts on the tangent components through
\begin{equation}
\boldsymbol Q=\begin{pmatrix}
\cos(2\pi/3)&-\sin(2\pi/3)\\
\sin(2\pi/3)&\cos(2\pi/3)
\end{pmatrix}.
\end{equation}
Covariance gives $\boldsymbol Q\bm v=\bm v$ and $\boldsymbol M\boldsymbol Q=\boldsymbol Q\boldsymbol M$.  Since $\boldsymbol Q$ has no nonzero invariant vector in the tangent plane, $\bm v=0$; a field parallel to $\bm n_1$ cannot generate a first-order tangent response.  Solving the commutation relation gives
\begin{equation}
\boldsymbol M=a\boldsymbol I_2+b\boldsymbol J,
\qquad
\boldsymbol J=\begin{pmatrix}0&-1\\1&0\end{pmatrix}.
\label{eq:c3_commutant_supp}
\end{equation}
Thus the two allowed maps are the tangent-plane identity and its in-plane $90^\circ$ rotation.  In three-dimensional notation they are $\boldsymbol\Pi_1=\boldsymbol I-\bm n_1\bm n_1^{\mathsf T}$ and $[\bm n_1]_\times$, where $[\bm n_1]_\times\bm v=\bm n_1\times\bm v$.  Indeed, $\boldsymbol\Pi_1\bB$ and $\bm n_1\times\bB$ are orthogonal and have the same norm, so they span the local tangent plane whenever the transverse field is nonzero.  Transitivity of $T_h$ generates the maps on the other three sublattices with the same two coefficients.  Therefore the complete linear response is
\begin{equation}
\delta\bm e_i^{(1)}=
a\boldsymbol\Pi_i\bB+b\,\bm n_i\times\bB.
\label{eq:two_mode_canting_supp}
\end{equation}
The coefficients $a$ and $b$ are response amplitudes to be eliminated, not material parameters.  Direct substitution into Eq.~\eqref{eq:complete_tangent_free_energy_supp} gives
\begin{equation}
\begin{aligned}
\frac{F^{(2)}}{B^2}={}&
-\frac49\left(7D_x-7D_y-8J-6A
-7\gamma_x-\gamma_{xy}-7\gamma_y\right)a^2\\
&-\frac{8\sqrt3}{9}
\left(2D_x+2D_y-\gamma_x+\gamma_y\right)ab\\
&-\frac43\left(D_x-D_y-2A
-3\gamma_x-3\gamma_{xy}-3\gamma_y\right)b^2
-\frac83ha.
\end{aligned}
\label{eq:two_mode_quadratic_energy_supp}
\end{equation}
The equations $\partial F^{(2)}/\partial a=\partial F^{(2)}/\partial b=0$ are
\begin{equation}
\begin{pmatrix}
7D_x-7D_y-8J-6A-7\gamma_x-\gamma_{xy}-7\gamma_y
&\sqrt3(2D_x+2D_y-\gamma_x+\gamma_y)\\
\sqrt3(2D_x+2D_y-\gamma_x+\gamma_y)
&3(D_x-D_y-2A-3\gamma_x-3\gamma_{xy}-3\gamma_y)
\end{pmatrix}
\begin{pmatrix}a\\b\end{pmatrix}
=-\begin{pmatrix}3h\\0\end{pmatrix}.
\label{eq:two_mode_stationarity_supp}
\end{equation}
Its determinant, after removing the common factor of three associated with the second row, is
\begin{equation}
\begin{aligned}
\Delta={}&
\left(D_x-D_y-2A-3\gamma_x-3\gamma_{xy}-3\gamma_y\right)\\
&\times\left(7D_x-7D_y-8J-6A
-7\gamma_x-\gamma_{xy}-7\gamma_y\right)\\
&-\left(2D_x+2D_y-\gamma_x+\gamma_y\right)^2.
\end{aligned}
\label{eq:two_mode_determinant_supp}
\end{equation}
The solution is
\begin{equation}
\begin{aligned}
a={}&-\frac{3h}{\Delta}
\left(D_x-D_y-2A-3\gamma_x-3\gamma_{xy}-3\gamma_y\right),\\
b={}&\frac{\sqrt3h}{\Delta}
\left(2D_x+2D_y-\gamma_x+\gamma_y\right).
\end{aligned}
\label{eq:two_mode_solution_supp}
\end{equation}
The matrix in Eq.~\eqref{eq:two_mode_stationarity_supp} is minus a positive factor times the physical quadratic stiffness in the $(a,b)$ subspace.  Its nonsingularity makes the driven response well defined, but it does not by itself establish metastability; canonical AIAO is metastable only if all eight eigenvalues of the full tangent Hessian $K$ are positive.

Equation~\eqref{eq:general_cubic_energy_supp} now reduces to
\begin{equation}
\begin{aligned}
F_\eta^{(3)}=\frac{4\eta}{9}(a^2+b^2)
\bigl\{&2\sqrt3(D_x-D_y+4J-\gamma_x+2\gamma_{xy}-\gamma_y)a\\
&-6(D_x+D_y+\gamma_x-\gamma_y)b-3\sqrt3h\bigr\}
B_xB_yB_z.
\end{aligned}
\label{eq:two_mode_cubic_energy_supp}
\end{equation}
Substituting Eq.~\eqref{eq:two_mode_solution_supp} yields $F_\eta^{(3)}=-\eta\lambda B_xB_yB_z$.  Hence $\chi_{\eta,xyz}^{(2)}=\eta\lambda$, with the same value for all six index permutations.  The fully explicit seven-parameter coefficient is
\begin{equation}
\begin{aligned}
\lambda={}&\frac{12\sqrt3h^3}{\Delta^3}
\Bigl\{3(D_x-D_y-2A-3\gamma_x-3\gamma_{xy}-3\gamma_y)^2\\
&\qquad +(2D_x+2D_y-\gamma_x+\gamma_y)^2\Bigr\}\\
&\times\Bigl\{3D_x^2-6D_xD_y-8D_xA
-10D_x\gamma_x-8D_x\gamma_{xy}-14D_x\gamma_y\\
&\qquad +3D_y^2+8D_yA+14D_y\gamma_x
+8D_y\gamma_{xy}+10D_y\gamma_y\\
&\qquad +4A^2+12A\gamma_x+4A\gamma_{xy}
+12A\gamma_y\\
&\qquad +8\gamma_x^2+6\gamma_x\gamma_{xy}+20\gamma_x\gamma_y
-3\gamma_{xy}^2+6\gamma_{xy}\gamma_y+8\gamma_y^2\Bigr\}.
\end{aligned}
\label{eq:explicit_lambda_supp}
\end{equation}
Apart from the determinant abbreviation $\Delta$, every quantity in Eq.~\eqref{eq:explicit_lambda_supp} is one of the seven coefficients of Eq.~\eqref{eq:seven_parameter_th_model_supp} or the physical Zeeman scale $h=\mu_Bm$.

\subsubsection{Tetrahedral identities.}

The canonical axes in Eq.~\eqref{eq:canonical_aiao_axes_supp} obey
\begin{equation}
\sum_{i=1}^{4}\bm n_i=0,
\qquad
\sum_{i=1}^{4}\bm n_i\bm n_i^{\mathsf T}=\frac43\boldsymbol I,
\qquad
\sum_{i=1}^{4}(\bB\cdot\bm n_i)^3=\frac{8}{\sqrt3}B_xB_yB_z.
\label{eq:tetrahedral_identities_supp}
\end{equation}
The first identity in Eq.~\eqref{eq:tetrahedral_identities_supp} cancels the linear Zeeman energy of the compensated AIAO reference, the second fixes the transverse field response, and the third produces the cubic angular dependence.

\begingroup
\makeatletter
\let\@FMN@list\@empty
\makeatother
\let\originalLabel\label
\renewcommand{\label}[1]{\originalLabel{SMBIB:#1}}

\input{supplement.bbl}
\endgroup

%% file: tables/tr_mpg_character_longtable.tex
\begingroup
\ifdefined\SUPPLEMENTREDLINE\color{red}\fi
\footnotesize
\newcommand{\mpg}[1]{\ensuremath{\mathrm{#1}}}
\setlength{\tabcolsep}{2.2pt}
\setlength{\LTcapwidth}{0.96\linewidth}
\setlength{\LTleft}{0pt}
\setlength{\LTright}{0pt}
\renewcommand{\arraystretch}{1.10}
\begin{longtable}{@{}lllll@{}}
\caption{
Complete channel-resolved classification of pure uniform-field domain selectors.  The 105 rows list the ordered-state subgroup \(\mathcal M\), its gray parent \(\mathcal G=\mathcal M+\mathcal M\mathcal T\), the lowest allowed odd rank \(n_{\min}\), the corresponding magnetic class used in the main text, and a basis of the leading field polynomials.  All \(B_x,B_y,B_z\) are components in an orthonormal Cartesian frame; for trigonal and hexagonal settings, \(x\) is chosen along the first conventional basal axis and \(z\) along the principal axis.  Magnetic-point-group symbols follow the BCS MPOINT convention~\cite{SM:BilbaoCrystallographicServer,SM:Aroyo2006BCS,SM:Aroyo2006BCSII,SM:BorovikRomanov2013,SM:Elcoro2021MagTQC,SM:Xu2020MagTopological}.  Semicolon-separated expressions form a basis.  Repeated subgroup symbols correspond to distinct parent-group embeddings.  A dash denotes exact pure-field silence at every odd order because \(\mathcal{PT}\in\mathcal M\).
}
\label{tab:mpg-character-scan}\\
\hline\hline
\begin{minipage}[t]{2.25cm}Subgroup \(\mathcal M\)\end{minipage} & \begin{minipage}[t]{2.25cm}Parent \(\mathcal G\)\end{minipage} & \begin{minipage}[t]{0.85cm}\(n_{\min}\)\end{minipage} & \begin{minipage}[t]{2.0cm}Class\end{minipage} & \begin{minipage}[t]{8.2cm}Basis of \(\phi_i^{n_{\min}}(\bm B)\)\end{minipage}\\
\hline
\endfirsthead
\hline\hline
\begin{minipage}[t]{2.25cm}Subgroup \(\mathcal M\)\end{minipage} & \begin{minipage}[t]{2.25cm}Parent \(\mathcal G\)\end{minipage} & \begin{minipage}[t]{0.85cm}\(n_{\min}\)\end{minipage} & \begin{minipage}[t]{2.0cm}Class\end{minipage} & \begin{minipage}[t]{8.2cm}Basis of \(\phi_i^{n_{\min}}(\bm B)\)\end{minipage}\\
\hline
\endhead
\hline
\multicolumn{5}{r}{continued on next page}\\
\endfoot
\hline\hline
\endlastfoot
\begin{minipage}[t]{2.25cm}\raggedright \mpg{\bar{1}}\end{minipage} & \begin{minipage}[t]{2.25cm}\raggedright \mpg{\bar{1}1'}\end{minipage} & \begin{minipage}[t]{0.85cm}\centering 1\end{minipage} & \begin{minipage}[t]{2.0cm}\raggedright FM\end{minipage} & \begin{minipage}[t]{8.2cm}\raggedright \(B_x; B_y; B_z\)\end{minipage} \\
\begin{minipage}[t]{2.25cm}\raggedright \mpg{\bar{1}'}\end{minipage} & \begin{minipage}[t]{2.25cm}\raggedright \mpg{\bar{1}1'}\end{minipage} & \begin{minipage}[t]{0.85cm}\centering --\end{minipage} & \begin{minipage}[t]{2.0cm}\raggedright \(\mathcal{PT}\)-AFM\end{minipage} & \begin{minipage}[t]{8.2cm}\raggedright --\end{minipage} \\
\begin{minipage}[t]{2.25cm}\raggedright \mpg{\bar{3}}\end{minipage} & \begin{minipage}[t]{2.25cm}\raggedright \mpg{\bar{3}1'}\end{minipage} & \begin{minipage}[t]{0.85cm}\centering 1\end{minipage} & \begin{minipage}[t]{2.0cm}\raggedright FM\end{minipage} & \begin{minipage}[t]{8.2cm}\raggedright \(B_z\)\end{minipage} \\
\begin{minipage}[t]{2.25cm}\raggedright \mpg{\bar{3}'}\end{minipage} & \begin{minipage}[t]{2.25cm}\raggedright \mpg{\bar{3}1'}\end{minipage} & \begin{minipage}[t]{0.85cm}\centering --\end{minipage} & \begin{minipage}[t]{2.0cm}\raggedright \(\mathcal{PT}\)-AFM\end{minipage} & \begin{minipage}[t]{8.2cm}\raggedright --\end{minipage} \\
\begin{minipage}[t]{2.25cm}\raggedright \mpg{\bar{3}m'}\end{minipage} & \begin{minipage}[t]{2.25cm}\raggedright \mpg{\bar{3}m1'}\end{minipage} & \begin{minipage}[t]{0.85cm}\centering 1\end{minipage} & \begin{minipage}[t]{2.0cm}\raggedright FM\end{minipage} & \begin{minipage}[t]{8.2cm}\raggedright \(B_z\)\end{minipage} \\
\begin{minipage}[t]{2.25cm}\raggedright \mpg{\bar{3}m}\end{minipage} & \begin{minipage}[t]{2.25cm}\raggedright \mpg{\bar{3}m1'}\end{minipage} & \begin{minipage}[t]{0.85cm}\centering 3\end{minipage} & \begin{minipage}[t]{2.0cm}\raggedright AM or nc-AFM\end{minipage} & \begin{minipage}[t]{8.2cm}\raggedright \(B_x\,(B_x^2\,-\,3\,B_y^2)\)\end{minipage} \\
\begin{minipage}[t]{2.25cm}\raggedright \mpg{\bar{3}'m'}\end{minipage} & \begin{minipage}[t]{2.25cm}\raggedright \mpg{\bar{3}m1'}\end{minipage} & \begin{minipage}[t]{0.85cm}\centering --\end{minipage} & \begin{minipage}[t]{2.0cm}\raggedright \(\mathcal{PT}\)-AFM\end{minipage} & \begin{minipage}[t]{8.2cm}\raggedright --\end{minipage} \\
\begin{minipage}[t]{2.25cm}\raggedright \mpg{\bar{3}'m}\end{minipage} & \begin{minipage}[t]{2.25cm}\raggedright \mpg{\bar{3}m1'}\end{minipage} & \begin{minipage}[t]{0.85cm}\centering --\end{minipage} & \begin{minipage}[t]{2.0cm}\raggedright \(\mathcal{PT}\)-AFM\end{minipage} & \begin{minipage}[t]{8.2cm}\raggedright --\end{minipage} \\
\begin{minipage}[t]{2.25cm}\raggedright \mpg{\bar{4}}\end{minipage} & \begin{minipage}[t]{2.25cm}\raggedright \mpg{\bar{4}1'}\end{minipage} & \begin{minipage}[t]{0.85cm}\centering 1\end{minipage} & \begin{minipage}[t]{2.0cm}\raggedright FM\end{minipage} & \begin{minipage}[t]{8.2cm}\raggedright \(B_z\)\end{minipage} \\
\begin{minipage}[t]{2.25cm}\raggedright \mpg{\bar{4}'}\end{minipage} & \begin{minipage}[t]{2.25cm}\raggedright \mpg{\bar{4}1'}\end{minipage} & \begin{minipage}[t]{0.85cm}\centering 3\end{minipage} & \begin{minipage}[t]{2.0cm}\raggedright AM or nc-AFM\end{minipage} & \begin{minipage}[t]{8.2cm}\raggedright \(B_x\,B_y\,B_z; B_z\,(B_x\,-\,B_y)\,(B_x\,+\,B_y)\)\end{minipage} \\
\begin{minipage}[t]{2.25cm}\raggedright \mpg{\bar{4}2'm'}\end{minipage} & \begin{minipage}[t]{2.25cm}\raggedright \mpg{\bar{4}2m1'}\end{minipage} & \begin{minipage}[t]{0.85cm}\centering 1\end{minipage} & \begin{minipage}[t]{2.0cm}\raggedright FM\end{minipage} & \begin{minipage}[t]{8.2cm}\raggedright \(B_z\)\end{minipage} \\
\begin{minipage}[t]{2.25cm}\raggedright \mpg{\bar{4}'2m'}\end{minipage} & \begin{minipage}[t]{2.25cm}\raggedright \mpg{\bar{4}2m1'}\end{minipage} & \begin{minipage}[t]{0.85cm}\centering 3\end{minipage} & \begin{minipage}[t]{2.0cm}\raggedright AM or nc-AFM\end{minipage} & \begin{minipage}[t]{8.2cm}\raggedright \(B_x\,B_y\,B_z\)\end{minipage} \\
\begin{minipage}[t]{2.25cm}\raggedright \mpg{\bar{4}'2'm}\end{minipage} & \begin{minipage}[t]{2.25cm}\raggedright \mpg{\bar{4}2m1'}\end{minipage} & \begin{minipage}[t]{0.85cm}\centering 3\end{minipage} & \begin{minipage}[t]{2.0cm}\raggedright AM or nc-AFM\end{minipage} & \begin{minipage}[t]{8.2cm}\raggedright \(B_z\,(B_x\,-\,B_y)\,(B_x\,+\,B_y)\)\end{minipage} \\
\begin{minipage}[t]{2.25cm}\raggedright \mpg{\bar{4}2m}\end{minipage} & \begin{minipage}[t]{2.25cm}\raggedright \mpg{\bar{4}2m1'}\end{minipage} & \begin{minipage}[t]{0.85cm}\centering 5\end{minipage} & \begin{minipage}[t]{2.0cm}\raggedright AM or nc-AFM\end{minipage} & \begin{minipage}[t]{8.2cm}\raggedright \(B_x\,B_y\,B_z\,(B_x\,-\,B_y)\,(B_x\,+\,B_y)\)\end{minipage} \\
\begin{minipage}[t]{2.25cm}\raggedright \mpg{\bar{4}'3m'}\end{minipage} & \begin{minipage}[t]{2.25cm}\raggedright \mpg{\bar{4}3m1'}\end{minipage} & \begin{minipage}[t]{0.85cm}\centering 3\end{minipage} & \begin{minipage}[t]{2.0cm}\raggedright AM or nc-AFM\end{minipage} & \begin{minipage}[t]{8.2cm}\raggedright \(B_x\,B_y\,B_z\)\end{minipage} \\
\begin{minipage}[t]{2.25cm}\raggedright \mpg{\bar{4}3m}\end{minipage} & \begin{minipage}[t]{2.25cm}\raggedright \mpg{\bar{4}3m1'}\end{minipage} & \begin{minipage}[t]{0.85cm}\centering 9\end{minipage} & \begin{minipage}[t]{2.0cm}\raggedright AM or nc-AFM\end{minipage} & \begin{minipage}[t]{8.2cm}\raggedright \(B_x\,B_y\,B_z\,(B_x\,-\,B_y)\,(B_x\,+\,B_y)\,(B_x\,-\,B_z)\,(B_x\,+\,B_z)\,(B_y\,-\,B_z)\,(B_y\,+\,B_z)\)\end{minipage} \\
\begin{minipage}[t]{2.25cm}\raggedright \mpg{\bar{6}}\end{minipage} & \begin{minipage}[t]{2.25cm}\raggedright \mpg{\bar{6}1'}\end{minipage} & \begin{minipage}[t]{0.85cm}\centering 1\end{minipage} & \begin{minipage}[t]{2.0cm}\raggedright FM\end{minipage} & \begin{minipage}[t]{8.2cm}\raggedright \(B_z\)\end{minipage} \\
\begin{minipage}[t]{2.25cm}\raggedright \mpg{\bar{6}'}\end{minipage} & \begin{minipage}[t]{2.25cm}\raggedright \mpg{\bar{6}1'}\end{minipage} & \begin{minipage}[t]{0.85cm}\centering 3\end{minipage} & \begin{minipage}[t]{2.0cm}\raggedright AM or nc-AFM\end{minipage} & \begin{minipage}[t]{8.2cm}\raggedright \(B_x\,(B_x^2\,-\,3\,B_y^2); B_y\,(3\,B_x^2\,-\,B_y^2)\)\end{minipage} \\
\begin{minipage}[t]{2.25cm}\raggedright \mpg{\bar{6}m'2'}\end{minipage} & \begin{minipage}[t]{2.25cm}\raggedright \mpg{\bar{6}m21'}\end{minipage} & \begin{minipage}[t]{0.85cm}\centering 1\end{minipage} & \begin{minipage}[t]{2.0cm}\raggedright FM\end{minipage} & \begin{minipage}[t]{8.2cm}\raggedright \(B_z\)\end{minipage} \\
\begin{minipage}[t]{2.25cm}\raggedright \mpg{\bar{6}'m'2}\end{minipage} & \begin{minipage}[t]{2.25cm}\raggedright \mpg{\bar{6}m21'}\end{minipage} & \begin{minipage}[t]{0.85cm}\centering 3\end{minipage} & \begin{minipage}[t]{2.0cm}\raggedright AM or nc-AFM\end{minipage} & \begin{minipage}[t]{8.2cm}\raggedright \(B_y\,(3\,B_x^2\,-\,B_y^2)\)\end{minipage} \\
\begin{minipage}[t]{2.25cm}\raggedright \mpg{\bar{6}'m2'}\end{minipage} & \begin{minipage}[t]{2.25cm}\raggedright \mpg{\bar{6}m21'}\end{minipage} & \begin{minipage}[t]{0.85cm}\centering 3\end{minipage} & \begin{minipage}[t]{2.0cm}\raggedright AM or nc-AFM\end{minipage} & \begin{minipage}[t]{8.2cm}\raggedright \(B_x\,(B_x^2\,-\,3\,B_y^2)\)\end{minipage} \\
\begin{minipage}[t]{2.25cm}\raggedright \mpg{\bar{6}m2}\end{minipage} & \begin{minipage}[t]{2.25cm}\raggedright \mpg{\bar{6}m21'}\end{minipage} & \begin{minipage}[t]{0.85cm}\centering 7\end{minipage} & \begin{minipage}[t]{2.0cm}\raggedright AM or nc-AFM\end{minipage} & \begin{minipage}[t]{8.2cm}\raggedright \(B_x\,B_y\,B_z\,(B_x^2\,-\,3\,B_y^2)\,(3\,B_x^2\,-\,B_y^2)\)\end{minipage} \\
\begin{minipage}[t]{2.25cm}\raggedright \mpg{1}\end{minipage} & \begin{minipage}[t]{2.25cm}\raggedright \mpg{11'}\end{minipage} & \begin{minipage}[t]{0.85cm}\centering 1\end{minipage} & \begin{minipage}[t]{2.0cm}\raggedright FM\end{minipage} & \begin{minipage}[t]{8.2cm}\raggedright \(B_x; B_y; B_z\)\end{minipage} \\
\begin{minipage}[t]{2.25cm}\raggedright \mpg{2/m}\end{minipage} & \begin{minipage}[t]{2.25cm}\raggedright \mpg{2/m1'}\end{minipage} & \begin{minipage}[t]{0.85cm}\centering 1\end{minipage} & \begin{minipage}[t]{2.0cm}\raggedright FM\end{minipage} & \begin{minipage}[t]{8.2cm}\raggedright \(B_y\)\end{minipage} \\
\begin{minipage}[t]{2.25cm}\raggedright \mpg{2'/m'}\end{minipage} & \begin{minipage}[t]{2.25cm}\raggedright \mpg{2/m1'}\end{minipage} & \begin{minipage}[t]{0.85cm}\centering 1\end{minipage} & \begin{minipage}[t]{2.0cm}\raggedright FM\end{minipage} & \begin{minipage}[t]{8.2cm}\raggedright \(B_x; B_z\)\end{minipage} \\
\begin{minipage}[t]{2.25cm}\raggedright \mpg{2/m'}\end{minipage} & \begin{minipage}[t]{2.25cm}\raggedright \mpg{2/m1'}\end{minipage} & \begin{minipage}[t]{0.85cm}\centering --\end{minipage} & \begin{minipage}[t]{2.0cm}\raggedright \(\mathcal{PT}\)-AFM\end{minipage} & \begin{minipage}[t]{8.2cm}\raggedright --\end{minipage} \\
\begin{minipage}[t]{2.25cm}\raggedright \mpg{2'/m}\end{minipage} & \begin{minipage}[t]{2.25cm}\raggedright \mpg{2/m1'}\end{minipage} & \begin{minipage}[t]{0.85cm}\centering --\end{minipage} & \begin{minipage}[t]{2.0cm}\raggedright \(\mathcal{PT}\)-AFM\end{minipage} & \begin{minipage}[t]{8.2cm}\raggedright --\end{minipage} \\
\begin{minipage}[t]{2.25cm}\raggedright \mpg{2}\end{minipage} & \begin{minipage}[t]{2.25cm}\raggedright \mpg{21'}\end{minipage} & \begin{minipage}[t]{0.85cm}\centering 1\end{minipage} & \begin{minipage}[t]{2.0cm}\raggedright FM\end{minipage} & \begin{minipage}[t]{8.2cm}\raggedright \(B_y\)\end{minipage} \\
\begin{minipage}[t]{2.25cm}\raggedright \mpg{2'}\end{minipage} & \begin{minipage}[t]{2.25cm}\raggedright \mpg{21'}\end{minipage} & \begin{minipage}[t]{0.85cm}\centering 1\end{minipage} & \begin{minipage}[t]{2.0cm}\raggedright FM\end{minipage} & \begin{minipage}[t]{8.2cm}\raggedright \(B_x; B_z\)\end{minipage} \\
\begin{minipage}[t]{2.25cm}\raggedright \mpg{22'2'}\end{minipage} & \begin{minipage}[t]{2.25cm}\raggedright \mpg{2221'}\end{minipage} & \begin{minipage}[t]{0.85cm}\centering 1\end{minipage} & \begin{minipage}[t]{2.0cm}\raggedright FM\end{minipage} & \begin{minipage}[t]{8.2cm}\raggedright \(B_x\)\end{minipage} \\
\begin{minipage}[t]{2.25cm}\raggedright \mpg{2'2'2}\end{minipage} & \begin{minipage}[t]{2.25cm}\raggedright \mpg{2221'}\end{minipage} & \begin{minipage}[t]{0.85cm}\centering 1\end{minipage} & \begin{minipage}[t]{2.0cm}\raggedright FM\end{minipage} & \begin{minipage}[t]{8.2cm}\raggedright \(B_y\)\end{minipage} \\
\begin{minipage}[t]{2.25cm}\raggedright \mpg{2'2'2}\end{minipage} & \begin{minipage}[t]{2.25cm}\raggedright \mpg{2221'}\end{minipage} & \begin{minipage}[t]{0.85cm}\centering 1\end{minipage} & \begin{minipage}[t]{2.0cm}\raggedright FM\end{minipage} & \begin{minipage}[t]{8.2cm}\raggedright \(B_z\)\end{minipage} \\
\begin{minipage}[t]{2.25cm}\raggedright \mpg{222}\end{minipage} & \begin{minipage}[t]{2.25cm}\raggedright \mpg{2221'}\end{minipage} & \begin{minipage}[t]{0.85cm}\centering 3\end{minipage} & \begin{minipage}[t]{2.0cm}\raggedright AM or nc-AFM\end{minipage} & \begin{minipage}[t]{8.2cm}\raggedright \(B_x\,B_y\,B_z\)\end{minipage} \\
\begin{minipage}[t]{2.25cm}\raggedright \mpg{23}\end{minipage} & \begin{minipage}[t]{2.25cm}\raggedright \mpg{231'}\end{minipage} & \begin{minipage}[t]{0.85cm}\centering 3\end{minipage} & \begin{minipage}[t]{2.0cm}\raggedright AM or nc-AFM\end{minipage} & \begin{minipage}[t]{8.2cm}\raggedright \(B_x\,B_y\,B_z\)\end{minipage} \\
\begin{minipage}[t]{2.25cm}\raggedright \mpg{3}\end{minipage} & \begin{minipage}[t]{2.25cm}\raggedright \mpg{31'}\end{minipage} & \begin{minipage}[t]{0.85cm}\centering 1\end{minipage} & \begin{minipage}[t]{2.0cm}\raggedright FM\end{minipage} & \begin{minipage}[t]{8.2cm}\raggedright \(B_z\)\end{minipage} \\
\begin{minipage}[t]{2.25cm}\raggedright \mpg{32'}\end{minipage} & \begin{minipage}[t]{2.25cm}\raggedright \mpg{321'}\end{minipage} & \begin{minipage}[t]{0.85cm}\centering 1\end{minipage} & \begin{minipage}[t]{2.0cm}\raggedright FM\end{minipage} & \begin{minipage}[t]{8.2cm}\raggedright \(B_z\)\end{minipage} \\
\begin{minipage}[t]{2.25cm}\raggedright \mpg{32}\end{minipage} & \begin{minipage}[t]{2.25cm}\raggedright \mpg{321'}\end{minipage} & \begin{minipage}[t]{0.85cm}\centering 3\end{minipage} & \begin{minipage}[t]{2.0cm}\raggedright AM or nc-AFM\end{minipage} & \begin{minipage}[t]{8.2cm}\raggedright \(B_x\,(B_x^2\,-\,3\,B_y^2)\)\end{minipage} \\
\begin{minipage}[t]{2.25cm}\raggedright \mpg{3m'}\end{minipage} & \begin{minipage}[t]{2.25cm}\raggedright \mpg{3m1'}\end{minipage} & \begin{minipage}[t]{0.85cm}\centering 1\end{minipage} & \begin{minipage}[t]{2.0cm}\raggedright FM\end{minipage} & \begin{minipage}[t]{8.2cm}\raggedright \(B_z\)\end{minipage} \\
\begin{minipage}[t]{2.25cm}\raggedright \mpg{3m}\end{minipage} & \begin{minipage}[t]{2.25cm}\raggedright \mpg{3m1'}\end{minipage} & \begin{minipage}[t]{0.85cm}\centering 3\end{minipage} & \begin{minipage}[t]{2.0cm}\raggedright AM or nc-AFM\end{minipage} & \begin{minipage}[t]{8.2cm}\raggedright \(B_x\,(B_x^2\,-\,3\,B_y^2)\)\end{minipage} \\
\begin{minipage}[t]{2.25cm}\raggedright \mpg{4/m}\end{minipage} & \begin{minipage}[t]{2.25cm}\raggedright \mpg{4/m1'}\end{minipage} & \begin{minipage}[t]{0.85cm}\centering 1\end{minipage} & \begin{minipage}[t]{2.0cm}\raggedright FM\end{minipage} & \begin{minipage}[t]{8.2cm}\raggedright \(B_z\)\end{minipage} \\
\begin{minipage}[t]{2.25cm}\raggedright \mpg{4'/m}\end{minipage} & \begin{minipage}[t]{2.25cm}\raggedright \mpg{4/m1'}\end{minipage} & \begin{minipage}[t]{0.85cm}\centering 3\end{minipage} & \begin{minipage}[t]{2.0cm}\raggedright AM or nc-AFM\end{minipage} & \begin{minipage}[t]{8.2cm}\raggedright \(B_x\,B_y\,B_z; B_z\,(B_x\,-\,B_y)\,(B_x\,+\,B_y)\)\end{minipage} \\
\begin{minipage}[t]{2.25cm}\raggedright \mpg{4/m'}\end{minipage} & \begin{minipage}[t]{2.25cm}\raggedright \mpg{4/m1'}\end{minipage} & \begin{minipage}[t]{0.85cm}\centering --\end{minipage} & \begin{minipage}[t]{2.0cm}\raggedright \(\mathcal{PT}\)-AFM\end{minipage} & \begin{minipage}[t]{8.2cm}\raggedright --\end{minipage} \\
\begin{minipage}[t]{2.25cm}\raggedright \mpg{4'/m'}\end{minipage} & \begin{minipage}[t]{2.25cm}\raggedright \mpg{4/m1'}\end{minipage} & \begin{minipage}[t]{0.85cm}\centering --\end{minipage} & \begin{minipage}[t]{2.0cm}\raggedright \(\mathcal{PT}\)-AFM\end{minipage} & \begin{minipage}[t]{8.2cm}\raggedright --\end{minipage} \\
\begin{minipage}[t]{2.25cm}\raggedright \mpg{4/mm'm'}\end{minipage} & \begin{minipage}[t]{2.25cm}\raggedright \mpg{4/mmm1'}\end{minipage} & \begin{minipage}[t]{0.85cm}\centering 1\end{minipage} & \begin{minipage}[t]{2.0cm}\raggedright FM\end{minipage} & \begin{minipage}[t]{8.2cm}\raggedright \(B_z\)\end{minipage} \\
\begin{minipage}[t]{2.25cm}\raggedright \mpg{4'/mmm'}\end{minipage} & \begin{minipage}[t]{2.25cm}\raggedright \mpg{4/mmm1'}\end{minipage} & \begin{minipage}[t]{0.85cm}\centering 3\end{minipage} & \begin{minipage}[t]{2.0cm}\raggedright AM or nc-AFM\end{minipage} & \begin{minipage}[t]{8.2cm}\raggedright \(B_x\,B_y\,B_z\)\end{minipage} \\
\begin{minipage}[t]{2.25cm}\raggedright \mpg{4'/mm'm}\end{minipage} & \begin{minipage}[t]{2.25cm}\raggedright \mpg{4/mmm1'}\end{minipage} & \begin{minipage}[t]{0.85cm}\centering 3\end{minipage} & \begin{minipage}[t]{2.0cm}\raggedright AM or nc-AFM\end{minipage} & \begin{minipage}[t]{8.2cm}\raggedright \(B_z\,(B_x\,-\,B_y)\,(B_x\,+\,B_y)\)\end{minipage} \\
\begin{minipage}[t]{2.25cm}\raggedright \mpg{4/mmm}\end{minipage} & \begin{minipage}[t]{2.25cm}\raggedright \mpg{4/mmm1'}\end{minipage} & \begin{minipage}[t]{0.85cm}\centering 5\end{minipage} & \begin{minipage}[t]{2.0cm}\raggedright AM or nc-AFM\end{minipage} & \begin{minipage}[t]{8.2cm}\raggedright \(B_x\,B_y\,B_z\,(B_x\,-\,B_y)\,(B_x\,+\,B_y)\)\end{minipage} \\
\begin{minipage}[t]{2.25cm}\raggedright \mpg{4/m'm'm'}\end{minipage} & \begin{minipage}[t]{2.25cm}\raggedright \mpg{4/mmm1'}\end{minipage} & \begin{minipage}[t]{0.85cm}\centering --\end{minipage} & \begin{minipage}[t]{2.0cm}\raggedright \(\mathcal{PT}\)-AFM\end{minipage} & \begin{minipage}[t]{8.2cm}\raggedright --\end{minipage} \\
\begin{minipage}[t]{2.25cm}\raggedright \mpg{4/m'mm}\end{minipage} & \begin{minipage}[t]{2.25cm}\raggedright \mpg{4/mmm1'}\end{minipage} & \begin{minipage}[t]{0.85cm}\centering --\end{minipage} & \begin{minipage}[t]{2.0cm}\raggedright \(\mathcal{PT}\)-AFM\end{minipage} & \begin{minipage}[t]{8.2cm}\raggedright --\end{minipage} \\
\begin{minipage}[t]{2.25cm}\raggedright \mpg{4'/m'm'm}\end{minipage} & \begin{minipage}[t]{2.25cm}\raggedright \mpg{4/mmm1'}\end{minipage} & \begin{minipage}[t]{0.85cm}\centering --\end{minipage} & \begin{minipage}[t]{2.0cm}\raggedright \(\mathcal{PT}\)-AFM\end{minipage} & \begin{minipage}[t]{8.2cm}\raggedright --\end{minipage} \\
\begin{minipage}[t]{2.25cm}\raggedright \mpg{4'/m'mm'}\end{minipage} & \begin{minipage}[t]{2.25cm}\raggedright \mpg{4/mmm1'}\end{minipage} & \begin{minipage}[t]{0.85cm}\centering --\end{minipage} & \begin{minipage}[t]{2.0cm}\raggedright \(\mathcal{PT}\)-AFM\end{minipage} & \begin{minipage}[t]{8.2cm}\raggedright --\end{minipage} \\
\begin{minipage}[t]{2.25cm}\raggedright \mpg{4}\end{minipage} & \begin{minipage}[t]{2.25cm}\raggedright \mpg{41'}\end{minipage} & \begin{minipage}[t]{0.85cm}\centering 1\end{minipage} & \begin{minipage}[t]{2.0cm}\raggedright FM\end{minipage} & \begin{minipage}[t]{8.2cm}\raggedright \(B_z\)\end{minipage} \\
\begin{minipage}[t]{2.25cm}\raggedright \mpg{4'}\end{minipage} & \begin{minipage}[t]{2.25cm}\raggedright \mpg{41'}\end{minipage} & \begin{minipage}[t]{0.85cm}\centering 3\end{minipage} & \begin{minipage}[t]{2.0cm}\raggedright AM or nc-AFM\end{minipage} & \begin{minipage}[t]{8.2cm}\raggedright \(B_x\,B_y\,B_z; B_z\,(B_x\,-\,B_y)\,(B_x\,+\,B_y)\)\end{minipage} \\
\begin{minipage}[t]{2.25cm}\raggedright \mpg{42'2'}\end{minipage} & \begin{minipage}[t]{2.25cm}\raggedright \mpg{4221'}\end{minipage} & \begin{minipage}[t]{0.85cm}\centering 1\end{minipage} & \begin{minipage}[t]{2.0cm}\raggedright FM\end{minipage} & \begin{minipage}[t]{8.2cm}\raggedright \(B_z\)\end{minipage} \\
\begin{minipage}[t]{2.25cm}\raggedright \mpg{4'22'}\end{minipage} & \begin{minipage}[t]{2.25cm}\raggedright \mpg{4221'}\end{minipage} & \begin{minipage}[t]{0.85cm}\centering 3\end{minipage} & \begin{minipage}[t]{2.0cm}\raggedright AM or nc-AFM\end{minipage} & \begin{minipage}[t]{8.2cm}\raggedright \(B_x\,B_y\,B_z\)\end{minipage} \\
\begin{minipage}[t]{2.25cm}\raggedright \mpg{4'2'2}\end{minipage} & \begin{minipage}[t]{2.25cm}\raggedright \mpg{4221'}\end{minipage} & \begin{minipage}[t]{0.85cm}\centering 3\end{minipage} & \begin{minipage}[t]{2.0cm}\raggedright AM or nc-AFM\end{minipage} & \begin{minipage}[t]{8.2cm}\raggedright \(B_z\,(B_x\,-\,B_y)\,(B_x\,+\,B_y)\)\end{minipage} \\
\begin{minipage}[t]{2.25cm}\raggedright \mpg{422}\end{minipage} & \begin{minipage}[t]{2.25cm}\raggedright \mpg{4221'}\end{minipage} & \begin{minipage}[t]{0.85cm}\centering 5\end{minipage} & \begin{minipage}[t]{2.0cm}\raggedright AM or nc-AFM\end{minipage} & \begin{minipage}[t]{8.2cm}\raggedright \(B_x\,B_y\,B_z\,(B_x\,-\,B_y)\,(B_x\,+\,B_y)\)\end{minipage} \\
\begin{minipage}[t]{2.25cm}\raggedright \mpg{4'32'}\end{minipage} & \begin{minipage}[t]{2.25cm}\raggedright \mpg{4321'}\end{minipage} & \begin{minipage}[t]{0.85cm}\centering 3\end{minipage} & \begin{minipage}[t]{2.0cm}\raggedright AM or nc-AFM\end{minipage} & \begin{minipage}[t]{8.2cm}\raggedright \(B_x\,B_y\,B_z\)\end{minipage} \\
\begin{minipage}[t]{2.25cm}\raggedright \mpg{432}\end{minipage} & \begin{minipage}[t]{2.25cm}\raggedright \mpg{4321'}\end{minipage} & \begin{minipage}[t]{0.85cm}\centering 9\end{minipage} & \begin{minipage}[t]{2.0cm}\raggedright AM or nc-AFM\end{minipage} & \begin{minipage}[t]{8.2cm}\raggedright \(B_x\,B_y\,B_z\,(B_x\,-\,B_y)\,(B_x\,+\,B_y)\,(B_x\,-\,B_z)\,(B_x\,+\,B_z)\,(B_y\,-\,B_z)\,(B_y\,+\,B_z)\)\end{minipage} \\
\begin{minipage}[t]{2.25cm}\raggedright \mpg{4m'm'}\end{minipage} & \begin{minipage}[t]{2.25cm}\raggedright \mpg{4mm1'}\end{minipage} & \begin{minipage}[t]{0.85cm}\centering 1\end{minipage} & \begin{minipage}[t]{2.0cm}\raggedright FM\end{minipage} & \begin{minipage}[t]{8.2cm}\raggedright \(B_z\)\end{minipage} \\
\begin{minipage}[t]{2.25cm}\raggedright \mpg{4'mm'}\end{minipage} & \begin{minipage}[t]{2.25cm}\raggedright \mpg{4mm1'}\end{minipage} & \begin{minipage}[t]{0.85cm}\centering 3\end{minipage} & \begin{minipage}[t]{2.0cm}\raggedright AM or nc-AFM\end{minipage} & \begin{minipage}[t]{8.2cm}\raggedright \(B_x\,B_y\,B_z\)\end{minipage} \\
\begin{minipage}[t]{2.25cm}\raggedright \mpg{4'm'm}\end{minipage} & \begin{minipage}[t]{2.25cm}\raggedright \mpg{4mm1'}\end{minipage} & \begin{minipage}[t]{0.85cm}\centering 3\end{minipage} & \begin{minipage}[t]{2.0cm}\raggedright AM or nc-AFM\end{minipage} & \begin{minipage}[t]{8.2cm}\raggedright \(B_z\,(B_x\,-\,B_y)\,(B_x\,+\,B_y)\)\end{minipage} \\
\begin{minipage}[t]{2.25cm}\raggedright \mpg{4mm}\end{minipage} & \begin{minipage}[t]{2.25cm}\raggedright \mpg{4mm1'}\end{minipage} & \begin{minipage}[t]{0.85cm}\centering 5\end{minipage} & \begin{minipage}[t]{2.0cm}\raggedright AM or nc-AFM\end{minipage} & \begin{minipage}[t]{8.2cm}\raggedright \(B_x\,B_y\,B_z\,(B_x\,-\,B_y)\,(B_x\,+\,B_y)\)\end{minipage} \\
\begin{minipage}[t]{2.25cm}\raggedright \mpg{6/m}\end{minipage} & \begin{minipage}[t]{2.25cm}\raggedright \mpg{6/m1'}\end{minipage} & \begin{minipage}[t]{0.85cm}\centering 1\end{minipage} & \begin{minipage}[t]{2.0cm}\raggedright FM\end{minipage} & \begin{minipage}[t]{8.2cm}\raggedright \(B_z\)\end{minipage} \\
\begin{minipage}[t]{2.25cm}\raggedright \mpg{6'/m'}\end{minipage} & \begin{minipage}[t]{2.25cm}\raggedright \mpg{6/m1'}\end{minipage} & \begin{minipage}[t]{0.85cm}\centering 3\end{minipage} & \begin{minipage}[t]{2.0cm}\raggedright AM or nc-AFM\end{minipage} & \begin{minipage}[t]{8.2cm}\raggedright \(B_x\,(B_x^2\,-\,3\,B_y^2); B_y\,(3\,B_x^2\,-\,B_y^2)\)\end{minipage} \\
\begin{minipage}[t]{2.25cm}\raggedright \mpg{6/m'}\end{minipage} & \begin{minipage}[t]{2.25cm}\raggedright \mpg{6/m1'}\end{minipage} & \begin{minipage}[t]{0.85cm}\centering --\end{minipage} & \begin{minipage}[t]{2.0cm}\raggedright \(\mathcal{PT}\)-AFM\end{minipage} & \begin{minipage}[t]{8.2cm}\raggedright --\end{minipage} \\
\begin{minipage}[t]{2.25cm}\raggedright \mpg{6'/m}\end{minipage} & \begin{minipage}[t]{2.25cm}\raggedright \mpg{6/m1'}\end{minipage} & \begin{minipage}[t]{0.85cm}\centering --\end{minipage} & \begin{minipage}[t]{2.0cm}\raggedright \(\mathcal{PT}\)-AFM\end{minipage} & \begin{minipage}[t]{8.2cm}\raggedright --\end{minipage} \\
\begin{minipage}[t]{2.25cm}\raggedright \mpg{6/mm'm'}\end{minipage} & \begin{minipage}[t]{2.25cm}\raggedright \mpg{6/mmm1'}\end{minipage} & \begin{minipage}[t]{0.85cm}\centering 1\end{minipage} & \begin{minipage}[t]{2.0cm}\raggedright FM\end{minipage} & \begin{minipage}[t]{8.2cm}\raggedright \(B_z\)\end{minipage} \\
\begin{minipage}[t]{2.25cm}\raggedright \mpg{6'/m'm'm}\end{minipage} & \begin{minipage}[t]{2.25cm}\raggedright \mpg{6/mmm1'}\end{minipage} & \begin{minipage}[t]{0.85cm}\centering 3\end{minipage} & \begin{minipage}[t]{2.0cm}\raggedright AM or nc-AFM\end{minipage} & \begin{minipage}[t]{8.2cm}\raggedright \(B_y\,(3\,B_x^2\,-\,B_y^2)\)\end{minipage} \\
\begin{minipage}[t]{2.25cm}\raggedright \mpg{6'/m'mm'}\end{minipage} & \begin{minipage}[t]{2.25cm}\raggedright \mpg{6/mmm1'}\end{minipage} & \begin{minipage}[t]{0.85cm}\centering 3\end{minipage} & \begin{minipage}[t]{2.0cm}\raggedright AM or nc-AFM\end{minipage} & \begin{minipage}[t]{8.2cm}\raggedright \(B_x\,(B_x^2\,-\,3\,B_y^2)\)\end{minipage} \\
\begin{minipage}[t]{2.25cm}\raggedright \mpg{6/mmm}\end{minipage} & \begin{minipage}[t]{2.25cm}\raggedright \mpg{6/mmm1'}\end{minipage} & \begin{minipage}[t]{0.85cm}\centering 7\end{minipage} & \begin{minipage}[t]{2.0cm}\raggedright AM or nc-AFM\end{minipage} & \begin{minipage}[t]{8.2cm}\raggedright \(B_x\,B_y\,B_z\,(B_x^2\,-\,3\,B_y^2)\,(3\,B_x^2\,-\,B_y^2)\)\end{minipage} \\
\begin{minipage}[t]{2.25cm}\raggedright \mpg{6/m'm'm'}\end{minipage} & \begin{minipage}[t]{2.25cm}\raggedright \mpg{6/mmm1'}\end{minipage} & \begin{minipage}[t]{0.85cm}\centering --\end{minipage} & \begin{minipage}[t]{2.0cm}\raggedright \(\mathcal{PT}\)-AFM\end{minipage} & \begin{minipage}[t]{8.2cm}\raggedright --\end{minipage} \\
\begin{minipage}[t]{2.25cm}\raggedright \mpg{6/m'mm}\end{minipage} & \begin{minipage}[t]{2.25cm}\raggedright \mpg{6/mmm1'}\end{minipage} & \begin{minipage}[t]{0.85cm}\centering --\end{minipage} & \begin{minipage}[t]{2.0cm}\raggedright \(\mathcal{PT}\)-AFM\end{minipage} & \begin{minipage}[t]{8.2cm}\raggedright --\end{minipage} \\
\begin{minipage}[t]{2.25cm}\raggedright \mpg{6'/mmm'}\end{minipage} & \begin{minipage}[t]{2.25cm}\raggedright \mpg{6/mmm1'}\end{minipage} & \begin{minipage}[t]{0.85cm}\centering --\end{minipage} & \begin{minipage}[t]{2.0cm}\raggedright \(\mathcal{PT}\)-AFM\end{minipage} & \begin{minipage}[t]{8.2cm}\raggedright --\end{minipage} \\
\begin{minipage}[t]{2.25cm}\raggedright \mpg{6'/mm'm}\end{minipage} & \begin{minipage}[t]{2.25cm}\raggedright \mpg{6/mmm1'}\end{minipage} & \begin{minipage}[t]{0.85cm}\centering --\end{minipage} & \begin{minipage}[t]{2.0cm}\raggedright \(\mathcal{PT}\)-AFM\end{minipage} & \begin{minipage}[t]{8.2cm}\raggedright --\end{minipage} \\
\begin{minipage}[t]{2.25cm}\raggedright \mpg{6}\end{minipage} & \begin{minipage}[t]{2.25cm}\raggedright \mpg{61'}\end{minipage} & \begin{minipage}[t]{0.85cm}\centering 1\end{minipage} & \begin{minipage}[t]{2.0cm}\raggedright FM\end{minipage} & \begin{minipage}[t]{8.2cm}\raggedright \(B_z\)\end{minipage} \\
\begin{minipage}[t]{2.25cm}\raggedright \mpg{6'}\end{minipage} & \begin{minipage}[t]{2.25cm}\raggedright \mpg{61'}\end{minipage} & \begin{minipage}[t]{0.85cm}\centering 3\end{minipage} & \begin{minipage}[t]{2.0cm}\raggedright AM or nc-AFM\end{minipage} & \begin{minipage}[t]{8.2cm}\raggedright \(B_x\,(B_x^2\,-\,3\,B_y^2); B_y\,(3\,B_x^2\,-\,B_y^2)\)\end{minipage} \\
\begin{minipage}[t]{2.25cm}\raggedright \mpg{62'2'}\end{minipage} & \begin{minipage}[t]{2.25cm}\raggedright \mpg{6221'}\end{minipage} & \begin{minipage}[t]{0.85cm}\centering 1\end{minipage} & \begin{minipage}[t]{2.0cm}\raggedright FM\end{minipage} & \begin{minipage}[t]{8.2cm}\raggedright \(B_z\)\end{minipage} \\
\begin{minipage}[t]{2.25cm}\raggedright \mpg{6'2'2}\end{minipage} & \begin{minipage}[t]{2.25cm}\raggedright \mpg{6221'}\end{minipage} & \begin{minipage}[t]{0.85cm}\centering 3\end{minipage} & \begin{minipage}[t]{2.0cm}\raggedright AM or nc-AFM\end{minipage} & \begin{minipage}[t]{8.2cm}\raggedright \(B_y\,(3\,B_x^2\,-\,B_y^2)\)\end{minipage} \\
\begin{minipage}[t]{2.25cm}\raggedright \mpg{6'22'}\end{minipage} & \begin{minipage}[t]{2.25cm}\raggedright \mpg{6221'}\end{minipage} & \begin{minipage}[t]{0.85cm}\centering 3\end{minipage} & \begin{minipage}[t]{2.0cm}\raggedright AM or nc-AFM\end{minipage} & \begin{minipage}[t]{8.2cm}\raggedright \(B_x\,(B_x^2\,-\,3\,B_y^2)\)\end{minipage} \\
\begin{minipage}[t]{2.25cm}\raggedright \mpg{622}\end{minipage} & \begin{minipage}[t]{2.25cm}\raggedright \mpg{6221'}\end{minipage} & \begin{minipage}[t]{0.85cm}\centering 7\end{minipage} & \begin{minipage}[t]{2.0cm}\raggedright AM or nc-AFM\end{minipage} & \begin{minipage}[t]{8.2cm}\raggedright \(B_x\,B_y\,B_z\,(B_x^2\,-\,3\,B_y^2)\,(3\,B_x^2\,-\,B_y^2)\)\end{minipage} \\
\begin{minipage}[t]{2.25cm}\raggedright \mpg{6m'm'}\end{minipage} & \begin{minipage}[t]{2.25cm}\raggedright \mpg{6mm1'}\end{minipage} & \begin{minipage}[t]{0.85cm}\centering 1\end{minipage} & \begin{minipage}[t]{2.0cm}\raggedright FM\end{minipage} & \begin{minipage}[t]{8.2cm}\raggedright \(B_z\)\end{minipage} \\
\begin{minipage}[t]{2.25cm}\raggedright \mpg{6'mm'}\end{minipage} & \begin{minipage}[t]{2.25cm}\raggedright \mpg{6mm1'}\end{minipage} & \begin{minipage}[t]{0.85cm}\centering 3\end{minipage} & \begin{minipage}[t]{2.0cm}\raggedright AM or nc-AFM\end{minipage} & \begin{minipage}[t]{8.2cm}\raggedright \(B_x\,(B_x^2\,-\,3\,B_y^2)\)\end{minipage} \\
\begin{minipage}[t]{2.25cm}\raggedright \mpg{6'm'm}\end{minipage} & \begin{minipage}[t]{2.25cm}\raggedright \mpg{6mm1'}\end{minipage} & \begin{minipage}[t]{0.85cm}\centering 3\end{minipage} & \begin{minipage}[t]{2.0cm}\raggedright AM or nc-AFM\end{minipage} & \begin{minipage}[t]{8.2cm}\raggedright \(B_y\,(3\,B_x^2\,-\,B_y^2)\)\end{minipage} \\
\begin{minipage}[t]{2.25cm}\raggedright \mpg{6mm}\end{minipage} & \begin{minipage}[t]{2.25cm}\raggedright \mpg{6mm1'}\end{minipage} & \begin{minipage}[t]{0.85cm}\centering 7\end{minipage} & \begin{minipage}[t]{2.0cm}\raggedright AM or nc-AFM\end{minipage} & \begin{minipage}[t]{8.2cm}\raggedright \(B_x\,B_y\,B_z\,(B_x^2\,-\,3\,B_y^2)\,(3\,B_x^2\,-\,B_y^2)\)\end{minipage} \\
\begin{minipage}[t]{2.25cm}\raggedright \mpg{m\bar{3}}\end{minipage} & \begin{minipage}[t]{2.25cm}\raggedright \mpg{m\bar{3}1'}\end{minipage} & \begin{minipage}[t]{0.85cm}\centering 3\end{minipage} & \begin{minipage}[t]{2.0cm}\raggedright AM or nc-AFM\end{minipage} & \begin{minipage}[t]{8.2cm}\raggedright \(B_x\,B_y\,B_z\)\end{minipage} \\
\begin{minipage}[t]{2.25cm}\raggedright \mpg{m'\bar{3}'}\end{minipage} & \begin{minipage}[t]{2.25cm}\raggedright \mpg{m\bar{3}1'}\end{minipage} & \begin{minipage}[t]{0.85cm}\centering --\end{minipage} & \begin{minipage}[t]{2.0cm}\raggedright \(\mathcal{PT}\)-AFM\end{minipage} & \begin{minipage}[t]{8.2cm}\raggedright --\end{minipage} \\
\begin{minipage}[t]{2.25cm}\raggedright \mpg{m\bar{3}m'}\end{minipage} & \begin{minipage}[t]{2.25cm}\raggedright \mpg{m\bar{3}m1'}\end{minipage} & \begin{minipage}[t]{0.85cm}\centering 3\end{minipage} & \begin{minipage}[t]{2.0cm}\raggedright AM or nc-AFM\end{minipage} & \begin{minipage}[t]{8.2cm}\raggedright \(B_x\,B_y\,B_z\)\end{minipage} \\
\begin{minipage}[t]{2.25cm}\raggedright \mpg{m\bar{3}m}\end{minipage} & \begin{minipage}[t]{2.25cm}\raggedright \mpg{m\bar{3}m1'}\end{minipage} & \begin{minipage}[t]{0.85cm}\centering 9\end{minipage} & \begin{minipage}[t]{2.0cm}\raggedright AM or nc-AFM\end{minipage} & \begin{minipage}[t]{8.2cm}\raggedright \(B_x\,B_y\,B_z\,(B_x\,-\,B_y)\,(B_x\,+\,B_y)\,(B_x\,-\,B_z)\,(B_x\,+\,B_z)\,(B_y\,-\,B_z)\,(B_y\,+\,B_z)\)\end{minipage} \\
\begin{minipage}[t]{2.25cm}\raggedright \mpg{m'\bar{3}'m'}\end{minipage} & \begin{minipage}[t]{2.25cm}\raggedright \mpg{m\bar{3}m1'}\end{minipage} & \begin{minipage}[t]{0.85cm}\centering --\end{minipage} & \begin{minipage}[t]{2.0cm}\raggedright \(\mathcal{PT}\)-AFM\end{minipage} & \begin{minipage}[t]{8.2cm}\raggedright --\end{minipage} \\
\begin{minipage}[t]{2.25cm}\raggedright \mpg{m'\bar{3}'m}\end{minipage} & \begin{minipage}[t]{2.25cm}\raggedright \mpg{m\bar{3}m1'}\end{minipage} & \begin{minipage}[t]{0.85cm}\centering --\end{minipage} & \begin{minipage}[t]{2.0cm}\raggedright \(\mathcal{PT}\)-AFM\end{minipage} & \begin{minipage}[t]{8.2cm}\raggedright --\end{minipage} \\
\begin{minipage}[t]{2.25cm}\raggedright \mpg{m}\end{minipage} & \begin{minipage}[t]{2.25cm}\raggedright \mpg{m1'}\end{minipage} & \begin{minipage}[t]{0.85cm}\centering 1\end{minipage} & \begin{minipage}[t]{2.0cm}\raggedright FM\end{minipage} & \begin{minipage}[t]{8.2cm}\raggedright \(B_y\)\end{minipage} \\
\begin{minipage}[t]{2.25cm}\raggedright \mpg{m'}\end{minipage} & \begin{minipage}[t]{2.25cm}\raggedright \mpg{m1'}\end{minipage} & \begin{minipage}[t]{0.85cm}\centering 1\end{minipage} & \begin{minipage}[t]{2.0cm}\raggedright FM\end{minipage} & \begin{minipage}[t]{8.2cm}\raggedright \(B_x; B_z\)\end{minipage} \\
\begin{minipage}[t]{2.25cm}\raggedright \mpg{m'm'2}\end{minipage} & \begin{minipage}[t]{2.25cm}\raggedright \mpg{mm21'}\end{minipage} & \begin{minipage}[t]{0.85cm}\centering 1\end{minipage} & \begin{minipage}[t]{2.0cm}\raggedright FM\end{minipage} & \begin{minipage}[t]{8.2cm}\raggedright \(B_z\)\end{minipage} \\
\begin{minipage}[t]{2.25cm}\raggedright \mpg{m'm2'}\end{minipage} & \begin{minipage}[t]{2.25cm}\raggedright \mpg{mm21'}\end{minipage} & \begin{minipage}[t]{0.85cm}\centering 1\end{minipage} & \begin{minipage}[t]{2.0cm}\raggedright FM\end{minipage} & \begin{minipage}[t]{8.2cm}\raggedright \(B_y\)\end{minipage} \\
\begin{minipage}[t]{2.25cm}\raggedright \mpg{mm'2'}\end{minipage} & \begin{minipage}[t]{2.25cm}\raggedright \mpg{mm21'}\end{minipage} & \begin{minipage}[t]{0.85cm}\centering 1\end{minipage} & \begin{minipage}[t]{2.0cm}\raggedright FM\end{minipage} & \begin{minipage}[t]{8.2cm}\raggedright \(B_x\)\end{minipage} \\
\begin{minipage}[t]{2.25cm}\raggedright \mpg{mm2}\end{minipage} & \begin{minipage}[t]{2.25cm}\raggedright \mpg{mm21'}\end{minipage} & \begin{minipage}[t]{0.85cm}\centering 3\end{minipage} & \begin{minipage}[t]{2.0cm}\raggedright AM or nc-AFM\end{minipage} & \begin{minipage}[t]{8.2cm}\raggedright \(B_x\,B_y\,B_z\)\end{minipage} \\
\begin{minipage}[t]{2.25cm}\raggedright \mpg{mm'm'}\end{minipage} & \begin{minipage}[t]{2.25cm}\raggedright \mpg{mmm1'}\end{minipage} & \begin{minipage}[t]{0.85cm}\centering 1\end{minipage} & \begin{minipage}[t]{2.0cm}\raggedright FM\end{minipage} & \begin{minipage}[t]{8.2cm}\raggedright \(B_x\)\end{minipage} \\
\begin{minipage}[t]{2.25cm}\raggedright \mpg{m'mm'}\end{minipage} & \begin{minipage}[t]{2.25cm}\raggedright \mpg{mmm1'}\end{minipage} & \begin{minipage}[t]{0.85cm}\centering 1\end{minipage} & \begin{minipage}[t]{2.0cm}\raggedright FM\end{minipage} & \begin{minipage}[t]{8.2cm}\raggedright \(B_y\)\end{minipage} \\
\begin{minipage}[t]{2.25cm}\raggedright \mpg{m'm'm}\end{minipage} & \begin{minipage}[t]{2.25cm}\raggedright \mpg{mmm1'}\end{minipage} & \begin{minipage}[t]{0.85cm}\centering 1\end{minipage} & \begin{minipage}[t]{2.0cm}\raggedright FM\end{minipage} & \begin{minipage}[t]{8.2cm}\raggedright \(B_z\)\end{minipage} \\
\begin{minipage}[t]{2.25cm}\raggedright \mpg{mmm}\end{minipage} & \begin{minipage}[t]{2.25cm}\raggedright \mpg{mmm1'}\end{minipage} & \begin{minipage}[t]{0.85cm}\centering 3\end{minipage} & \begin{minipage}[t]{2.0cm}\raggedright AM or nc-AFM\end{minipage} & \begin{minipage}[t]{8.2cm}\raggedright \(B_x\,B_y\,B_z\)\end{minipage} \\
\begin{minipage}[t]{2.25cm}\raggedright \mpg{m'm'm'}\end{minipage} & \begin{minipage}[t]{2.25cm}\raggedright \mpg{mmm1'}\end{minipage} & \begin{minipage}[t]{0.85cm}\centering --\end{minipage} & \begin{minipage}[t]{2.0cm}\raggedright \(\mathcal{PT}\)-AFM\end{minipage} & \begin{minipage}[t]{8.2cm}\raggedright --\end{minipage} \\
\begin{minipage}[t]{2.25cm}\raggedright \mpg{m'mm}\end{minipage} & \begin{minipage}[t]{2.25cm}\raggedright \mpg{mmm1'}\end{minipage} & \begin{minipage}[t]{0.85cm}\centering --\end{minipage} & \begin{minipage}[t]{2.0cm}\raggedright \(\mathcal{PT}\)-AFM\end{minipage} & \begin{minipage}[t]{8.2cm}\raggedright --\end{minipage} \\
\begin{minipage}[t]{2.25cm}\raggedright \mpg{mm'm}\end{minipage} & \begin{minipage}[t]{2.25cm}\raggedright \mpg{mmm1'}\end{minipage} & \begin{minipage}[t]{0.85cm}\centering --\end{minipage} & \begin{minipage}[t]{2.0cm}\raggedright \(\mathcal{PT}\)-AFM\end{minipage} & \begin{minipage}[t]{8.2cm}\raggedright --\end{minipage} \\
\begin{minipage}[t]{2.25cm}\raggedright \mpg{mmm'}\end{minipage} & \begin{minipage}[t]{2.25cm}\raggedright \mpg{mmm1'}\end{minipage} & \begin{minipage}[t]{0.85cm}\centering --\end{minipage} & \begin{minipage}[t]{2.0cm}\raggedright \(\mathcal{PT}\)-AFM\end{minipage} & \begin{minipage}[t]{8.2cm}\raggedright --\end{minipage} \\
\end{longtable}
\endgroup